\documentclass[physrev,reprint,showkeys,amsfonts,amssymb,amsmath]{revtex4-1}
\usepackage[T1]{fontenc}
\usepackage[utf8]{inputenc}
\usepackage[spanish,english]{babel}
\usepackage{hyperref}
\hypersetup{hidelinks}
\usepackage{physics}
\usepackage{booktabs}
\usepackage{enumerate}

\newcommand{\MeV}{\text{MeV}}

\newcommand{\fm}{\text{fm}}

\begin{document}

\title{Radiative decays in bottomonium beyond the long wave length approximation}
\author{R. Bruschini}
\email{roberto.bruschini@ific.uv.es}
\affiliation{\foreignlanguage{spanish}{Departamento de Física Teórica, IFIC, Universidad de Valencia-CSIC, E-46100 Burjassot (Valencia)}, Spain}
\author{P. González}
\email{pedro.gonzalez@uv.es}
\affiliation{\foreignlanguage{spanish}{Departamento de Física Teórica, IFIC, Universidad de Valencia-CSIC, E-46100 Burjassot (Valencia)}, Spain}


\begin{abstract}
We revisit the nonrelativistic quark model description of electromagnetic radiative decays in bottomonium. We show that  even for the simplest spectroscopic quark model  the calculated widths can be in good agreement with data once the experimental masses of bottomonium states and the photon energy are properly implemented in the calculation. For transitions involving the lower lying spectral states this implementation can be easily done via the Long Wave Length approximation. For transitions where this approximation does not apply we develop a new method of implementing the experimental energy dependencies.
\end{abstract}

\maketitle

\section{Introduction\label{SI}}

Electromagnetic radiative decays of hadrons provide useful information on the hadron structure. Their quark model description is based on the Elementary Emission Model (EEM) that assumes that the decay takes place through the emission of the photon by a quark (or antiquark) of the hadron, see for example \cite{LeY88}. As the electromagnetic transition operator is known, without any free parameter, radiative decays may be a powerful tool to discriminate among different spectroscopic hadron models.

In practice this discrimination may be rather difficult. Think, for example, of heavy quarkonium (bottomonium and charmonium) for which the nonrelativistic quark potential model is undoubtedly the more successful one in the spectral description of states below the open flavor meson-meson thresholds, see for instance \cite{Eic05} and references therein. (This is so even for the low lying charmonium states for which the calculated speed of the quark $Q$, or the antiquark $\overline{Q}$, given by $\frac{\lvert \vb*{p}_{Q}\rvert }{M_{Q}}$ where $\vb*{p}_{Q}$ $\left(M_{Q}\right)  $ is the three-momentum (mass) of the quark, can be about half of the speed of light.) Then, in order to build the electromagnetic transition operator for $I\rightarrow\gamma F$, where $I$ and $F$ are bottomonium or charmonium states, a nonrelativistic reduction of the well known point like quark photon interaction up to $\frac{\lvert \vb*{p}_{Q}\rvert }{M_{Q}}$ order is carried out. Moreover, for transitions where the wave length of the emitted photon is larger than the hadronic size scale of the process the operator is further simplified to the so called Long Wave Length Approximation (LWLA). Hence the comparison of calculated radiative widths to data may be testing not only the hadron structure model but also the hadron decay model approximation. This could be the reason why different spectroscopic quark models are successful (or fail) in the description of the same radiative decays \cite{Eic05,GI85,Seg16}.

\bigskip

In this article we center on bottomonium for which the nonrelativistic spectroscopic quark model as well as the nonrelativistic form of the electromagnetic transition operator can be reasonably taken for granted, and examine the requirements needed to get an accurate general description of radiative decays. We shall show that such a description may be attained, even from the simplest quark potential model wave functions reasonably fitting the spectroscopy, when the calculated mass differences between bottomonium states approximate the experimental ones. In the case that there is a discrepancy of tens of MeV at most, a good description is still feasible if the measured masses are properly implemented in the calculation.

\bigskip

The contents of the article are organized as follows. In Sec.~\ref{SII} we use the simplest spectroscopic (Cornell) potential model, yet incorporating the basic QCD ingredients for a physical description of bottomonium, for the calculation of the masses of the $S$ and $P$ spin triplet states far below open flavor thresholds. In Sec.~\ref{SIII} we recall the nonrelativistic form of the electromagnetic operator and focus on $S\longleftrightarrow P$ transitions between spin triplet states since these are quantitatively the more important ones and there are more data available. In Sec.~\ref{SIV} we take the LWLA that permits to factor out the final and initial state mass difference dependence in the operator. This allows us to implement the experimental masses in the calculation what turns out to be crucial for an accurate description of decays within the range of validity of the LWLA. In Sec.~\ref{SV} we pursue the mass difference factorization in the general case to get a good description of measured decays beyond the LWLA range of validity and to generate reliable predictions for not yet measured ones. Finally in Sec.~\ref{SVI} our main results and conclusions are summarized.

\section{Spectroscopic Quark Model \label{SII}}

The simplest non relativistic quark model physical description of bottomonium $\left(  b\overline{b}\right)  $ comes out from the hamiltonian
\begin{equation}
H_{C}=\frac{\vb*{p}^{2}}{M_{b}}+V_{C}\left(  r\right)
\label{hamc}
\end{equation}
where $V_{C}\left(  r\right)  $ is a Cornell like potential
\cite{Eic05,Eic80,Eic94}
\begin{equation}
V_{C}\left(  r\right)  =\sigma r-\frac{\zeta}{r}
\label{Cor}
\end{equation}
with $r$ standing for the $b-\overline{b}$ radial distance and $\sigma$ and $\zeta$ for the string tension and the chromoelectric coulomb strength parameters respectively. This static potential form has been justified from quenched lattice QCD calculations, see \cite{Bal01} and references therein. It should be kept in mind though that in the spirit of the nonrelativistic quark model calculations $\sigma$ and $\zeta$ are effective parameters through which some non considered corrections to the potential may be implicitly taken into account. Any different set of values of the parameters $\sigma,$ $\zeta$ and $M_{b}$ defines a different Cornell potential model. From now on we fix the Coulomb strength to $\zeta=100$ MeV fm corresponding to a strong quark-gluon coupling $\alpha_{s}=\frac{3\zeta}{4\hbar}\simeq0.38$ in agreement with the value derived from QCD from the hyperfine splitting of $1p$ states in bottomonium \cite{Ynd95}. As for $\sigma$ we expect, from lattice studies \cite{Bal01} a value around $900$ MeV/fm. Then, we choose it altogether with the quark mass, $M_{b}$, to get a reasonable fit, within a few tens of MeV, to the masses of $0^{-}\left(  1^{--}\right)  $ and $0^{+}\left(  J^{++}\right)$, $J=0,1,2,$ spin triplet states (let us recall that the neglected spin-spin contribution to the mass is three times smaller for triplet than for singlet states). More precisely, we define our model by (notice that this model does not contain any additive constant in the potential)
\begin{equation}
\begin{array}
[c]{c}
\sigma=850\,\MeV/\fm\\
\zeta=100\,\MeV\,\fm\\
M_{b}=4793\,\MeV
\end{array}
\label{parameters}
\end{equation}
from which a reasonable overall description of the spectral masses is obtained as shown in Table~\ref{Tabmassbbbar}.

\begin{table}
\centering
\begin{tabular}{ccccc}
\toprule
$J^{PC}$ & $
\begin{array}{c}
\text{Cornell}\\
nL\;\text{States}
\end{array}
$ & $
\begin{array}{c}
M_{Cor}\\
\text{MeV}
\end{array}
$ & $
\begin{array}
[c]{c}
M_{PDG}\\
\text{MeV}
\end{array}
$ & $
\begin{array}{c}
\langle r^{2}\rangle ^{\frac{1}{2}}\\
\text{fm}
\end{array}
$\\
\midrule
$1^{--}$ & $1S$ & $9459$ & $9460.30\pm0.26$ & $0.22$\\
& $2S$ & $10012$ & $10023.026\pm0.31$ & $0.51$\\
& $1S$ & $10157$ & $10163.7\pm1.4$ & \\
& $3S$ & $10342$ & $10355.2\pm0.5$ & $0.75$\\
& $2D$ & $10438$ &  & \\
& $4S$ & $10608$ & $10579.4\pm1.2$ & $0.96$\\
& $3D$ & $10682$ &  & \\
& $5S$ & $10841$ &  & $1.15$\\
&  &  & $10889.9_{-2.6}^{+3.2}$ & \\
& $4D$ & $10902$ &  & \\
\midrule
$0^{++}$ & $1P$ & $9920$ & $9859.44\pm0.42\pm0.31$ & $0.41$\\
$1^{++}$ & $1P$ & $9920$ & $9892.78\pm0.26\pm0.31$ & $0.41$\\
$2^{++}$ & $1P$ & $9920$ & $9912.21\pm0.26\pm0.31$ & $0.41$\\
\midrule
$0^{++}$ & $2P$ & $10259$ & $10232.5\pm0.4\pm0.5$ & $0.67$\\
$1^{++}$ & $2P$ & $10259$ & $10255.46\pm0.22\pm0.50$ & $0.67$\\
$2^{++}$ & $2P$ & $10259$ & $10268.65\pm0.22\pm0.50$ & $0.67$\\
\midrule
$0^{++}$ & $3P$ & $10531$ &  & $0.88$\\
$1^{++}$ & $3P$ & $10531$ & $10513.4\pm0.7$ & $0.88$\\
$2^{++}$ & $3P$ & $10531$ & $10524.0\pm0.8$ & $0.88$\\
\bottomrule
\end{tabular}
\caption{\label{Tabmassbbbar}Calculated $1^{--}$ and $J^{++}$ bottomonium masses, $M_{Cor}$,  far below their corresponding $S-$ wave open flavor meson-meson threshold (see for example \cite{Gon14} for a compilation of the values of these thresholds). The spectroscopic notation $nL$, where $n$ and $L$ are the radial and orbital angular momentum numbers respectively, has been used to characterize the $H_C$ eigenstates. Masses for experimental resonances, $M_{PDG}$, have been taken from \cite{PDG18}. For $p$ waves we quote separately the $np_{0}$, $np_{1}$ and $np_{2}$ states. The root mean square radii for the calculated states, $\left\langle r^{2}\right\rangle ^{\frac{1}{2}}$, are also reported.}
\end{table}

Some comments are in order. First, the low lying $1^{--}$ masses are well reproduced within $15$ MeV. The discrepancy between the calculated mass of the $4S$ state at $10608$ MeV and the experimental mass at $10579.4$ MeV may be indicating mixing of the $4S$ and $3D$ states. So, the measured resonance would have a dominant $4S$ component, whereas a not yet discovered resonance at about $10750$ MeV would have a dominant $3D$ component. Notice that $S-D$ mixing should be also present for the $5S$ and $4D$ states, apart from a possible additional mixing with the lowest hybrid state \cite{Bru19}. For higher $1^{--}$ states, not included in the table, the first $S-$ wave open flavor meson-meson threshold, $B\overline{B}_{1}$ at $11003$ MeV, may play an important role. 

Second, the calculated masses for $1P$ and $2P$ and $3P$ states differ from the corresponding measured $^{3}\!P_{2}$ masses by less than $10$ MeV. Therefore we may consider that our model fits reasonably well the $1^{--}$, $2^{++}$, and to a lesser extent $1^{++}$, spectroscopy.

Third, the calculated speed of the quark or antiquark is at most $0.3\,c$ what justifies the nonrelativistic form of the electromagnetic operator up to $\frac{\lvert \vb*{p}_{b}\rvert }{M_{b}}$ order we shall make use of.

\bigskip

Certainly potential corrections should be incorporated to the model for a more accurate description of the spectrum. We shall assume henceforth that for states far below (about $100$ MeV or more) their corresponding $S-$ wave open flavor meson-meson thresholds these corrections may be taken into account, at least to some extent, via first order perturbation theory. Then the model provides us with an appropriate set of bottomonium wave functions to be tested.

It is also worth to point out that although this model does not contain couple channel corrections it has proved to be useful as a starting point for the implicit incorporation (through the modification of the potential) of dominant spectroscopic threshold effects in bottomonium as well as in charmonium \cite{Gon14,Gon15,BrG19}. Alternatively, couple channel corrections have been explicitly implemented through unquenched quark models from a nonrelativistic \cite{Eic94,Eic04} or a semirelativistic \cite{Fer13,*Fer14_5,*Fer14_9} quark-antiquark Hamiltonian.

\section{Electromagnetic Decay Model \label{SIII}}

Let us consider the decay $I\rightarrow\gamma F$ where $I$ and $F$ are the initial and final bottomonium states respectively. In the rest frame of the decaying meson $I$ the total width is given by (we follow the PDG conventions, see \cite[p.~567]{PDG18})
\begin{equation}
\Gamma_{I\rightarrow\gamma F}=\frac{k_{0}}{8\pi M_{I}^{2}}\frac{1}{\left(2J_{I}+1\right)  }\sum_{\lambda=\pm1}\sum_{m_{I},m_{F}}\left\vert\mathcal{M}_{J_{F},m_{F},J_{I},m_{I}}^{\lambda}\right\vert ^{2}
\label{width}
\end{equation}
where $k_{0}$ is the energy of the photon and $M_{I},$ $J_{I}$ and $m_{I}$ stand for the mass of $I$, its total angular momentum and its third projection respectively$.$ The polarization of the photon is represented by $\lambda$ (as usual we choose the three-momentum of the photon in the $Z$ direction) and the transition matrix element by $\mathcal{M}_{J_{F},m_{F},J_{I},m_{I}}^{\lambda}$. This matrix element can be obtained from the interaction hamiltonian $\mathcal{H}_{int}$ as
\begin{multline}
\left(  2\pi\right)  ^{3}\delta^{(3)}\left(  \vb*{P}_{I}-\vb*{k}-\vb*{P}_{F}\right)  \mathcal{M}_{J_{F},m_{F},J_{I},m_{I}}^{\lambda}=\\
\sqrt{2M_{I}}\sqrt{2E_{F}}\sqrt{2k_{0}}\left\langle F\gamma\right\vert \mathcal{H}_{int}\left\vert I\right\rangle \label{amplitude1}
\end{multline}
where $\left(  E_{I},\vb*{P}_{I}\right)  =\left(M_{I},\vb*{0}\right)  ,$ $\left(  E_{F},\vb*{P}_{F}\right)  $ and $\left(  k_{0},\vb*{k}\right)$ are the meson and photon four-momenta.

\bigskip

In the Elementary Emission Decay Model the radiative transition $I\rightarrow\gamma F$ takes place through the emission of the photon by the quark or the antiquark of $I$. By proceeding to a nonrelativistic reduction of the interaction hamiltonian at the quark level (we use the radiation gauge so that the time component of the electromagnetic field vanishes, $A^{0}\left(\vb*{x}\right)  =0$) the operator to be sandwhiched between the meson states reads, see for example \cite{LeY88},
\begin{multline}
\left\langle \vb*{k},\lambda\right\vert \mathcal{H}_{I}\left\vert 0\right\rangle  =-\frac{1}{\sqrt{2k_{0}}}\sum_{\alpha=1,2}\frac{e_{\alpha}}{2M_{\alpha}}\\
\left(  e^{-i\vb*{k}\vdot\vb*{r}_{\alpha}}\vb*{p}_{\alpha}+\vb*{p}_{\alpha}e^{-i\vb*{k}\vdot\vb*{r}_{\alpha}}-i\vb*{\sigma}_{\alpha}\times\vb*{k}e^{-i\vb*{k}\vdot\vb*{r}_{\alpha}}\right) \vdot \left(  \vb*{\epsilon}_{\vb*{k}}^{\lambda}\right)^{\ast}
\label{transop}
\end{multline}
where the subindices $1$ and $2$ refer to quark $b$ and antiquark $\overline{b}$ respectively, $e_{1}$ $\left(  e_{2}\right)  $ is the $b$ ($\overline{b}$) electric charge, $e_{b}=-\frac{1}{3}\left\vert e\right\vert$, $\vb*{\epsilon}_{\vb*{k}}^{\lambda}$ stands for the photon polarization vector and $\vb*{k}$ is now a vector number, not an operator. Then, having into account the quantum numbers characterizing the initial and final states
\begin{equation}
\left\vert I\right\rangle =\left\vert \vb*{P}_{I},J_{I},m_{I},\left(  n_{I}L_{I}\right)  _{b\overline{b}},\left(  S_{I}\right)_{b\overline{b}}\right\rangle
\label{IN}
\end{equation}
\begin{equation}
\left\vert F\right\rangle =\left\vert \vb*{P}_{F},J_{F},m_{F},\left(  n_{F}L_{F}\right)  _{b\overline{b}},\left(  S_{F}\right)_{b\overline{b}}\right\rangle
\label{FIN}
\end{equation}
where $\left\vert \left(  nL\right)  _{b\overline{b}}\right\rangle $ stand for the $H_C$ eigenstates previously calculated, introducing center of mass
\begin{equation}
\vb*{R}=\frac{\vb*{r}_{1}+\vb*{r}_{2}}{2}, \quad\vb*{P}=\vb*{p}_{1}+\vb*{p}_{2}
\end{equation}
and relative
\begin{equation}
\vb*{r}=\vb*{r}_{1}-\vb*{r}_{2},\quad\vb*{p}=\frac{\vb*{p}_{1}-\vb*{p}_{2}}{2}
\end{equation}
operators and integrating over $\vb*{R},$ the center of mass spatial degrees of freedom, the transition matrix element can be written as
\begin{multline}
\mathcal{M}_{J_{F},m_{F},J_{I},m_{I}}^{\lambda}  =\sqrt{2M_{I}}\sqrt{2E_{F}}\sum_{\alpha=1,2}\frac{e_{\alpha}}{2M_{\alpha}}\\
\left\langle J_{F},m_{F},\left(  n_{F}L_{F}\right)  _{b\overline{b}},\left(  S_{F}\right)  _{b\overline{b}}\right\vert \\
 \overline{\mathcal{O}}_{\alpha}\left\vert J_{I},m_{I},\left(  n_{I}L_{I}\right)  _{b\overline{b}},\left(  S_{I}\right)  _{b\overline{b}}\right\rangle
\label{M}
\end{multline}
where
\begin{multline}
\overline{\mathcal{O}}_{\alpha} =\left(  \left(  -1\right)^{\alpha}\left(  e^{i(-1)^{\alpha}\left(  \frac{\vb*{k}\vdot\vb*{r}}{2}\right)  }\vb*{p}+\vb*{p}e^{i(-1)^{\alpha}\left(  \frac{\vb*{k}\vdot\vb*{r}}{2}\right)  }\right)  \right.\\
+i\vb*{\sigma}_{\alpha}\times\vb*{k}e^{i\left(-1\right)  ^{\alpha}\left(  \frac{\vb*{k}\vdot\vb*{r}}{2}\right)  }\\
\left.  -\left(  \frac{\vb*{P}_{I}+\vb*{P}_{F}}{2}\right)  e^{i(-1)^{\alpha}\left(  \frac{\vb*{k}\vdot\vb*{r}}{2}\right)  }\right)\vdot \left(  \vb*{\epsilon}_{\vb*{k}}^{\lambda}\right)^{\ast}
 \label{O}
\end{multline}
The first, second and third addends on the right hand side correspond to
electric, magnetic and convective terms respectively since they come from the
corresponding terms in the quark electromagnetic current entering in the
interaction hamiltonian.

\bigskip

For practical calculations we use
\begin{align}
\left[  (\vb*{p})_q ,e^{i\left(  -1\right)^{\alpha}\left(  \frac{\vb*{k}\vdot\vb*{r}}{2}\right)}\right]  &= \sum_{q'} \left[ (\vb*{p})_q, (\vb*{r})_{q'}\right]  \frac{\partial e^{i(-1)^{\alpha}\left(  \frac{\vb*{k}\vdot\vb*{r}}{2}\right)  }}{\partial (\vb*{r})_{q'}}\nonumber \\
&=(-1)^{\alpha}\frac{(\vb*{k})_q}{2}e^{i(-1)^{\alpha}\left(  \frac{\vb*{k}\vdot\vb*{r}}{2}\right)  }
\label{CONM}
\end{align}
or equivalently
\begin{equation}
\vb*{p}e^{i(-1)^{\alpha}\left(  \frac{\vb*{k}\vb*{r}}{2}\right)  }=e^{i(-1)^{\alpha}\left(\frac{\vb*{k}\vdot\vb*{r}}{2}\right)  }\vb*
{p}+(-1)^{\alpha}\frac{\vb*{k}}{2}e^{i\left(-1\right)  ^{\alpha}\left(  \frac{\vb*{k}\vdot\vb*{r}}
{2}\right)  }
\label{EQUIV}
\end{equation}
Then, by realizing that in the rest frame of the decaying meson $\vb*{P}_{I}=\vb*{0}$ and $\vb*{P}_{F}=-\vb*{k}$, where $\vb*{k}$ is in the $Z$ direction, one has $\vb*{P}_{F}\vdot\left(  \vb*{\epsilon}_{\vb*{k}}^{\lambda}\right)^{\ast}=\vb*{0}=\vb*{k}\vdot \left(  \vb*{\epsilon}_{\vb*{k}}^{\lambda}\right)^{\ast}$, and the operator $\overline{\mathcal{O}}_{\alpha}$ reduces to
\begin{equation}
\mathcal{O}_{\alpha}=\left(  e^{i(-1)^{\alpha}\left(\frac{\vb*{k}\vdot\vb*{r}}{2}\right)  }\left(  \left(-1\right)  ^{\alpha}2\vb*{p}+i\vb*{\sigma}_{\alpha}\times\vb*{k}\right)  \right) \vdot \left(  \vb*{\epsilon}_{\vb*{k}}^{\lambda}\right)  ^{\ast}
\label{Ored}
\end{equation}
or equivalently to
\begin{equation}
\mathcal{O}_{\alpha}^{\prime}=\left(  \left(  \left(-1\right)  ^{\alpha}2\vb*{p}+i\vb*{\sigma}_{\alpha}\times\vb*{k}\right)  e^{i(-1)^{\alpha}\left(\frac{\vb*{k}\vdot\vb*{r}}{2}\right)  }\right) \vdot \left(  \vb*{\epsilon}_{\vb*{k}}^{\lambda}\right)  ^{\ast}
\label{Oprimered}
\end{equation}
Detailed expressions for the direct calculation of elecric and magnetic amplitudes in configuration space for ${^{3}\!S_{1}}\rightarrow\gamma\,{^{3}\!P_{J}}$ and ${^{3}\!P_{J}}\rightarrow\gamma\,{^{3}\!S_{1}}$ transitions can be found in Appendices \ref{SVII} and \ref{SVIII}.

\bigskip

It is important to emphasize that the $\vb*{p}$ operator in \eqref{Ored} or \eqref{Oprimered} makes the matrix element on the r.h.s. of \eqref{M} to have a specific dependence on the $H_c$ eigenvalues for the initial and final states, see below. Indeed, the explicit extraction of this dependence will become essential for an accurate description of radiative decays.

\section{Long Wave Length Approximation (LWLA) \label{SIV}}

In the limit that the wave length of the emitted photon is sufficiently large as compared to the hadronic size scale of the process (we shall be more quantitative below) we can approximate $e^{i\left(  -1\right)^{\alpha}\left(  \frac{\vb*{k}\vdot\vb*{r}}{2}\right)}\simeq1$. This simplifies the transition operator to
\begin{equation}
\left(  \mathcal{O}_{\alpha}\right)  _{LWLA}=\left(  \left(  -1\right)^{\alpha}2\vb*{p}+i\vb*{\sigma}_{\alpha}\times\vb*{k}\right) \vdot \left(  \vb*{\epsilon}_{\vb*{k}}^{\lambda}\right)  ^{\ast}  =\left(  \mathcal{O}_{\alpha}^{\prime}\right)_{LWLA}
\label{OLWLA}
\end{equation}
Furthermore, using
\begin{equation}
\vb*{p}=-i\frac{M_{b}}{2}\left[  \vb*{r},H_{C}\right]
\label{conm}
\end{equation}
we get
\begin{multline}
\left(  \mathcal{M}_{J_{F},m_{F},J_{I},m_{I}}^{\lambda}\right)  _{LWLA}=\sqrt{2M_{I}}\sqrt{2E_{F}}\sum_{\alpha=1,2}\frac{e_{\alpha}}{2}\\
\left\langle J_{F},m_{F},\left(  n_{F}L_{F}\right)  _{b\overline{b}},\left(S_{F}\right)  _{b\overline{b}}\right\vert \\
(-1)^{\alpha}\left(  -i\right)  \left(  M_{I}-M_{F}\right)\vb*{r}+i\vb*{\sigma}_{\alpha}\times\vb*{k} \\
\left\vert J_{I},m_{I},\left(  n_{I}L_{I}\right)  _{b\overline{b}},\left(S_{I}\right)  _{b\overline{b}}\right\rangle \vdot \left(\vb*{\epsilon}_{\vb*{k}}^{\lambda}\right)^{\ast}
\label{LWLA}
\end{multline}
where we have substituted the difference between the $H_{C}$ eigenvalues for the initial and final states by their mass difference. Moreover, for values of $\left\vert \vb*{k}\right\vert =k_{0}$ such that $\frac{\vb*{k}^{2}}{2M_{F}}\ll M_{F}$ we can neglect the kinetic energy of the final meson and substitute $M_{I}-M_{F}\simeq k_{0}$ and $E_{F}\simeq M_{F}.$

\bigskip

It is very important to remark that in the LWLA:
\begin{enumerate}[i)]
\item the amplitude does not depend explicitly on the quark mass;
\item the mass dependence has been explicitly factored out.
\end{enumerate}
Therefore, if we implement the experimental masses in the calculation then the comparison of the calculated widths with data is directly testing the spectroscopic model wave functions (the underlying
assumption justifying this procedure is that the difference between the calculated masses and the experimental ones can be obtained in most cases from these wave functions by applying first order perturbation theory).

\bigskip

For radiative transitions like ${^{3}\!S_{1}}\rightarrow\gamma\,{^{3}\!P_{J}}$ and ${^{3}\!P_{J}}\rightarrow\gamma\,{^{3}\!S_{1}}$ , with $J=0,1,2,$ the magnetic term does not contribute, as one can easily check from \eqref{B3} when $\abs{\vb*{k}}\abs{\vb*{r}}\to 0$. Thus, in the LWLA these transitions are purely electric dipole $E1$ transitions. More precisely, using $\vb*{r}\vdot\left(\vb*{\epsilon}_{\vb*{k}}^{\lambda}\right)  ^{\ast}=\sqrt{\frac{4\pi}{3}}\left(  Y_{1}^{\lambda}\left(\widehat{r}\right)  \right)  ^{\ast}r$ and some angular momentum algebra we can write the amplitude as
\begin{widetext}
\begin{multline}
\left(  \mathcal{M}_{J_{F},m_{F},J_{I},m_{I}}^{\lambda}\right) _{LWLA}  =i\sqrt{2M_{I}}\sqrt{2E_{F}}e_{b}\left(-1\right)  ^{L_{I}}\sqrt{2L_{F}+1}C_{1,\text{ }J_{F},\text{ }J_{I}}^{\lambda,\text{ }m_{F},\text{ }m_{I}}\\
\left(\begin{array}{ccc}
L_{F} & 1 & L_{I}\\
0 & 0 & 0
\end{array}\right)
\left[\begin{array}{ccc}
1 & L_{F} & L_{I}\\
S_{F} & J_{I} & J_{F}
\end{array}\right]
\left(  M_{I}-M_{F}\right) \int_{0}^{\infty}\dd{r}\text{ }r^{2}\left(  R_{n_{F}L_{F}}\right)^{\ast}rR_{n_{I}L_{I}}
\label{ExpLWLA}
\end{multline}
\end{widetext}
where $R_{n_{I}L_{I}}$ $\left(  R_{n_{F}L_{F}}\right)  $ is the radial wave function of the initial (final) state, 
\begin{equation}
C_{1,\text{ }J_{F},\text{ }J_{I}}^{\lambda,\text{ }m_{F},\text{ }m_{I}}\equiv(-1)^{J_{F}-1-m_{I}}\sqrt{2J_{I}+1}
\left(\begin{array}{ccc}
1 & J_{F} & J_{I}\\
\lambda & m_{F} & -m_{I}
\end{array}\right)
\label{coefnew}
\end{equation}
with $\left(  {}\right)  $ standing for the $3j$ symbol, and
\begin{multline}
\left[\begin{array}{ccc}
1 & L_{F} & L_{I}\\
S_{F} & J_{I} & J_{F}
\end{array}\right]
\equiv(-1)^{1+L_{F}+S_{F}+J_{I}} \\
\sqrt{\left(2L_{I}+1\right)  \left(  2J_{F}+1\right)  }
\left\{\begin{array}{ccc}
1 & L_{F} & L_{I}\\
S_{F} & J_{I} & J_{F}
\end{array}\right\}
\label{newcoef}
\end{multline}
with $\left\{  {}\right\}  $ standing for the $6j$ symbol.

\bigskip

From \eqref{ExpLWLA} and \eqref{width} and using $M_{I}-M_{F}\simeq k_{0}$, $e_{b}=-\frac{1}{3}\left\vert e\right\vert$ and $\left\vert e\right\vert ^{2}=4\pi\hat\alpha,$ where $\hat\alpha\simeq\frac{1}{137}$ is the fine structure constant, the LWLA width reads
\begin{multline}
\Gamma_{LWLA}=\frac{4\hat\alpha k_{0}^{3}E_{F}}{27M_{I}}\left(  2L_{F}+1\right)
\left(\begin{array}{ccc}
L_{F} & 1 & L_{I}\\
0 & 0 & 0
\end{array}\right)^{2} \\
\left(  2L_{I}+1\right) \left(  2J_{F}+1\right)
\left\{\begin{array}{ccc}
1 & L_{F} & L_{I}\\
1 & J_{I} & J_{F}
\end{array}\right\}  ^{2} \\
\left\vert \int_{0}^{\infty}\dd{r}\text{ }r^{2}\left(  R_{n_{F}L_{F}}\right)  ^{\ast}rR_{n_{I}L_{I}}\right\vert ^{2}
\label{widthLWLA}
\end{multline}
which is just the standard expression for the dipole electric amplitude in the literature, see for instance \cite{Eic08}, if one takes into account that
\begin{equation}
\left(  2L_{F}+1\right)
\left(\begin{array}{ccc}
L_{F} & 1 & L_{I}\\
0 & 0 & 0
\end{array}\right)  ^{2}
\left(  2L_{I}+1\right)  =\max\left(  L_{I},L_{F}\right)
\label{ident}
\end{equation}

For the practical application of \eqref{ExpLWLA} the range of validity of the LWLA needs to be established. For this purpose we may reason that for values of $\left\vert \vb*{r}\right\vert \geq 2\left\langle r^{2}\right\rangle ^{1/2}_F$ , where $2\left\langle r^{2}\right\rangle ^{1/2}_F$ approximates the size of the final state (notice that the size of the initial state is always bigger), the radial wave function for this state almost vanishes giving a negligible contribution to the matrix element \eqref{ExpLWLA}. Hence for values of $\left\vert\vb*{k}\right\vert $ such that $\left\vert \vb*{k}\right\vert 2\left\langle r^{2}\right\rangle ^{1/2}_{F}<1$ we expect that the values of $\left\vert \vb*{r}\right\vert$ contributing dominantly to the matrix element satisfy $\left\vert\vb*{k}\right\vert \left\vert \vb*{r}\right\vert<\frac{1}{2}\Rightarrow e^{i(-1)^{\alpha}\left(  \frac{\vb*{k}\vdot\vb*{r}}{2}\right)  }\simeq1$. Hence we may adopt
\begin{equation}
\left\vert \vb*{k}\right\vert 2\left\langle r^{2}\right\rangle ^{1/2}_{F}<1 \label{criterium}
\end{equation}
as a criterion of validity of the LWLA. In Table~\ref{TabkvaluesSP} and Table~\ref{TabkvaluesPS} we list the experimental values, $\left\vert \vb*{k}\right\vert _{Exp}=\left(\frac{M_{I}^{2}-M_{F}^{2}}{2M_{I}}\right)  _{Exp},$ and the calculated values of $\left\vert \vb*{k}\right\vert _{Exp} 2\left\langle r^{2}\right\rangle ^{1/2}_{F}$ from our spectroscopic model for ${^{3}\!S_{1}}\rightarrow\gamma\,{^{3}\!P_{J}}$ and ${^{3}\!P_{J}}\rightarrow\gamma\,{^{3}\!S_{1}}$ transitions.

\begin{table}
\centering
\begin{tabular}{ccc}
\toprule
${^{3}\!S_{1}}\rightarrow\gamma\,{^{3}\!P_{J}}$ & $\left\vert \vb*
{k}\right\vert _{Exp}\text{(MeV)}$ & $\left\vert \vb*{k}\right\vert
_{Exp}\left(  2\left\langle r^{2}\right\rangle ^{\frac{1}{2}}\right)
_{^{3}\!P_{J}}$\\
\midrule
$\Upsilon\left(  2S\right)  \rightarrow\gamma\chi_{b_{0}}\left(  1P\right)  $
& $162.2$ & $0.67$\\
$\Upsilon\left(  2S\right)  \rightarrow\gamma\chi_{b_{1}}\left(  1P\right)  $
& $129.4$ & $0.54$\\
$\Upsilon\left(  2S\right)  \rightarrow\gamma\chi_{b_{2}}\left(  1P\right)  $
& $110.2$ & $0.46$\\
\midrule
$\Upsilon\left(  3S\right)  \rightarrow\gamma\chi_{b_{0}}\left(  2P\right)  $
& $122.3$ & $0.83$\\
$\Upsilon\left(  3S\right)  \rightarrow\gamma\chi_{b_{1}}\left(  2P\right)  $
& $99.5$ & $0.68$\\
$\Upsilon\left(  3S\right)  \rightarrow\gamma\chi_{b_{2}}\left(  2P\right)  $
& $86.6$ & $0.59$\\
$\Upsilon\left(  3S\right)  \rightarrow\gamma\chi_{b_{0}}\left(  1P\right)  $
& $484.1$ & $2.01$\\
$\Upsilon\left(  3S\right)  \rightarrow\gamma\chi_{b_{1}}\left(  1P\right)  $
& $451.7$ & $1.88$\\
$\Upsilon\left(  3S\right)  \rightarrow\gamma\chi_{b_{2}}\left(  1P\right)  $
& $433.5$ & $1.80$\\
\midrule
$\Upsilon\left(  4S\right)  \rightarrow\gamma\chi_{b_{1}}\left(  3P\right)  $
& $65.8$ & $0.59$\\
$\Upsilon\left(  4S\right)  \rightarrow\gamma\chi_{b_{2}}\left(  3P\right)  $
& $55.3$ & $0.47$\\
$\Upsilon\left(  4S\right)  \rightarrow\gamma\chi_{b_{0}}\left(  2P\right)  $
& $341.2$ & $2.32$\\
$\Upsilon\left(  4S\right)  \rightarrow\gamma\chi_{b_{1}}\left(  2P\right)  $
& $319.0$ & $2.17$\\
$\Upsilon\left(  4S\right)  \rightarrow\gamma\chi_{b_{2}}\left(  2P\right)  $
& $306.2$ & $2.08$\\
$\Upsilon\left(  4S\right)  \rightarrow\gamma\chi_{b_{0}}\left(  1P\right)  $
& $695.5$ & $2.89$\\
$\Upsilon\left(  4S\right)  \rightarrow\gamma\chi_{b_{1}}\left(  1P\right)  $
& $664.3$ & $2.76$\\
$\Upsilon\left(  4S\right)  \rightarrow\gamma\chi_{b_{2}}\left(  1P\right)  $
& $646.2$ & $2.69$\\
\midrule
$\Upsilon\left(  5S\right)  \rightarrow\gamma\chi_{b_{1}}\left(  3P\right)  $
& $370.0$ & $3.68$\\
$\Upsilon\left(  5S\right)  \rightarrow\gamma\chi_{b_{2}}\left(  3P\right)  $
& $359.8$ & $3.57$\\
$\Upsilon\left(  5S\right)  \rightarrow\gamma\chi_{b_{0}}\left(  2P\right)  $
& $637.6$ & $4.32$\\
$\Upsilon\left(  5S\right)  \rightarrow\gamma\chi_{b_{1}}\left(  2P\right)  $
& $616.0$ & $4.18$\\
$\Upsilon\left(  5S\right)  \rightarrow\gamma\chi_{b_{2}}\left(  2P\right)  $
& $603.5$ & $4.10$\\
$\Upsilon\left(  5S\right)  \rightarrow\gamma\chi_{b_{0}}\left(  1P\right)  $
& $981.7$ & $4.08$\\
$\Upsilon\left(  5S\right)  \rightarrow\gamma\chi_{b_{1}}\left(  1P\right)  $
& $951.5$ & $3.95$\\
$\Upsilon\left(  5S\right)  \rightarrow\gamma\chi_{b_{2}}\left(  1P\right)  $
& $933.8$ & $3.88$\\
\bottomrule
\end{tabular}
\caption{\label{TabkvaluesSP}Experimental values of the photon energy $\left\vert \vb*{k}\right\vert _{Exp}$ and calculated values of $\left\vert \vb*{k}\right\vert _{Exp}(  2\left\langle r^{2}\right\rangle ^{\frac{1}{2}})  _{^{3}\!P_{J}}$ from our model for ${^{3}\!S_{1}}\rightarrow\gamma\,{^{3}\!P_{J}}$ radiative transitions.}
\end{table}

\begin{table}
\centering
\begin{tabular}{ccc}
\toprule
${^{3}\!P_{J}}\rightarrow\gamma\,{^{3}\!S_{1}}$ & $\left\vert \vb*
{k}\right\vert _{Exp}\text{(MeV)}$ & $\left\vert \vb*{k}\right\vert
_{Exp}\left(  2\left\langle r^{2}\right\rangle ^{\frac{1}{2}}\right)
_{^{3}\!S_{1}}$\\
\midrule
$\chi_{b_{0}}\left(  1P\right)  \rightarrow\gamma\Upsilon\left(  1S\right)  $
& $390.9$ & $0.87$\\
$\chi_{b_{1}}\left(  1P\right)  \rightarrow\gamma\Upsilon\left(  1S\right)  $
& $423.5$ & $0.94$\\
$\chi_{b_{2}}\left(  1P\right)  \rightarrow\gamma\Upsilon\left(  1S\right)  $
& $441.7$ & $0.98$\\
\midrule
$\chi_{b_{0}}\left(  2P\right)  \rightarrow\gamma\Upsilon\left(  2S\right)  $
& $206.9$ & $1.07$\\
$\chi_{b_{1}}\left(  2P\right)  \rightarrow\gamma\Upsilon\left(  2S\right)  $
& $229.4$ & $1.19$\\
$\chi_{b_{2}}\left(  2P\right)  \rightarrow\gamma\Upsilon\left(  2S\right)  $
& $243.1$ & $1.26$\\
\midrule
$\chi_{b_{0}}\left(  2P\right)  \rightarrow\gamma\Upsilon\left(  1S\right)  $
& $742.9$ & $1.66$\\
$\chi_{b_{1}}\left(  2P\right)  \rightarrow\gamma\Upsilon\left(  1S\right)  $
& $764.2$ & $1.70$\\
$\chi_{b_{2}}\left(  2P\right)  \rightarrow\gamma\Upsilon\left(  1S\right)  $
& $777.1$ & $1.73$\\
\midrule
$\chi_{b_{1}}\left(  3P\right)  \rightarrow\gamma\Upsilon\left(  3S\right)  $
& $157.0$ & $1.19$\\
$\chi_{b_{2}}\left(  3P\right)  \rightarrow\gamma\Upsilon\left(  3S\right)  $
& $167.5$ & $1.27$\\
\midrule
$\chi_{b_{1}}\left(  3P\right)  \rightarrow\gamma\Upsilon\left(  2S\right)  $
& $478.7$ & $2.47$\\
$\chi_{b_{2}}\left(  3P\right)  \rightarrow\gamma\Upsilon\left(  2S\right)  $
& $488.8$ & $2.53$\\
\midrule
$\chi_{b_{1}}\left(  3P\right)  \rightarrow\gamma\Upsilon\left(  1S\right)  $
& $1000.4$ & $2.23$\\
$\chi_{b_{2}}\left(  3P\right)  \rightarrow\gamma\Upsilon\left(  1S\right)  $
& $1009.9$ & $2.25$\\
\bottomrule
\end{tabular}
\caption{\label{TabkvaluesPS}Experimental values of the photon energy $\left\vert \vb*{k}\right\vert _{Exp}$ and calculated values of $\left\vert \vb*{k}\right\vert _{Exp}\left(  2\left\langle r^{2}\right\rangle ^{\frac{1}{2}}\right)  _{^{3}\!S_{1}}$ from our model for ${^{3}\!P_{J}}\rightarrow\gamma\,{^{3}\!S_{1}}$ radiative transitions.}
\end{table}

We see that, according to our criterion, the LWLA can only be valid for $\Upsilon\left(  2S\right)  \rightarrow\gamma\chi_{b_J}\left(  1P\right)$, $\Upsilon\left(  3S\right)  \rightarrow\gamma\chi_{b_J}\left(  2P\right)$, $\Upsilon\left(  4S\right)  \rightarrow\gamma\chi_{b_J}\left(  3P\right)  $ and $\chi_{b_J}\left(  1P\right)  \rightarrow\gamma\Upsilon\left(  1S\right)  $. As a test we can compare the widths $\Gamma_{LWLA}^{\left(  The\right)  }$ obtained from \eqref{widthLWLA}, where the superindex $(The)$ means that they are obtained from the calculated spectral masses (and the calculated $\left\vert \vb*{k}\right\vert _{The}$ from them), with the corresponding widths $\Gamma_{p/M}^{\left(  The\right)  }$ obtained from \eqref{A4}, \eqref{A11} and \eqref{B3}, when the complete operator $\mathcal{O}_{\alpha}$ in \eqref{Ored} (for ${^{3}\!S_{1}}\rightarrow\gamma\,{^{3}\!P_{J}}$) and $\mathcal{O}_{\alpha}^{\prime}$ in \eqref{Oprimered} (for ${^{3}\!P_{J}}\rightarrow\gamma\,{^{3}\!S_{1}}$) are used. The results are shown in
Table~\ref{TabwidthsLWLA}, first and fifth columns respectively.

\begin{table*}
\centering
\begin{tabular}{cccccc}
\toprule
Radiative Decay & $
\begin{array}{c}
\Gamma_{LWLA}^{\left(  The\right)  }\\
\text{KeV}
\end{array}
$ & $
\begin{array}{c}
\Gamma_{LWLA}^{\left(  The-Exp\right)  }\\
\text{KeV}
\end{array}
$ & $
\begin{array}{c}
\Gamma_{Exp}^{PDG}\\
\text{KeV}
\end{array}
$ & $
\begin{array}{c}
\Gamma_{p/M}^{\left(  Mixed\right)  }\\
\text{KeV}
\end{array}
$ & $
\begin{array}{c}
\Gamma_{p/M}^{\left(  The\right)  }\\
\text{KeV}
\end{array}
$\\
\midrule
$\Upsilon\left(  2S\right)  \rightarrow\gamma\chi_{b_{0}}\left(  1P\right)  $
& $0.30$ & $1.61$ & $1.2\pm0.3$ & $0.5$ & $0.29$\\
$\Upsilon\left(  2S\right)  \rightarrow\gamma\chi_{b_{1}}\left(  1P\right)  $
& $0.89$ & $2.46$ & $2.2\pm0.3$ & $1.28$ & $0.91$\\
$\Upsilon\left(  2S\right)  \rightarrow\gamma\chi_{b_{2}}\left(  1P\right)  $
& $1.48$ & $2.55$ & $2.3\pm0.3$ & $1.82$ & $1.51$\\
\midrule
$\Upsilon\left(  3S\right)  \rightarrow\gamma\chi_{b_{0}}\left(  2P\right)  $
& $0.54$ & $1.72$ & $1.14\pm0.20$ & $0.77$ & $0.54$\\
$\Upsilon\left(  3S\right)  \rightarrow\gamma\chi_{b_{1}}\left(  2P\right)  $
& $1.63$ & $2.78$ & $2.6\pm0.5$ & $1.99$ & $1.66$\\
$\Upsilon\left(  3S\right)  \rightarrow\gamma\chi_{b_{2}}\left(  2P\right)  $
& $2.72$ & $3.03$ & $2.7\pm0.5$ & $2.87$ & $2.77$\\
\midrule
$\Upsilon\left(  4S\right)  \rightarrow\gamma\chi_{b_{0}}\left(  3P\right)  $
& $0.75$ & $1.06$ & & $0.83$ & $0.74$ \\
$\Upsilon\left(  4S\right)  \rightarrow\gamma\chi_{b_{1}}\left(  3P\right)  $
& $2.24$ & $1.47$ & & 1.97 & 2.27 \\
$\Upsilon\left(  4S\right)  \rightarrow\gamma\chi_{b_{2}}\left(  3P\right)  $
& $3.74$ & $1.37$ & & 2.70 & 3.79 \\
\midrule
$\chi_{b_{0}}\left(  1P\right)  \rightarrow\gamma\Upsilon\left(  1S\right)  $
& $34.58$ & $22.82$ & & $33.23$ & $38.58$\\
$\chi_{b_{1}}\left(  1P\right)  \rightarrow\gamma\Upsilon\left(  1S\right)  $
& $34.58$ & $28.81$ & & $32.05$ & $33.71$\\
$\chi_{b_{2}}\left(  1P\right)  \rightarrow\gamma\Upsilon\left(  1S\right)  $
& $34.58$ & $32.71$ & & $33.21$ & $33.73$\\
\bottomrule
\end{tabular}
\caption{\label{TabwidthsLWLA}Calculated widths to order $p/M$ as compared to data for $\Upsilon\left(  2S\right)  \rightarrow\gamma\chi_{b_J}\left(  1P\right)$, $\Upsilon\left(  3S\right)  \rightarrow\gamma\chi_{b_J}\left(  2P\right)$, $\Upsilon\left(  4S\right)  \rightarrow\gamma\chi_{b_J}\left(  3P\right)  $ and $\chi_{b_J}\left(  1P\right)  \rightarrow\gamma\Upsilon\left(  1S\right)  $. Our educated guess for the unknown $\chi_{b0}\left(  3P\right)  $ mass has been $10492$ MeV. Notation as follows. $\Gamma_{LWLA}^{\left(  The\right)  }$: width in the LWLA without any external input. $\Gamma_{LWLA}^{\left(The-Exp\right)  }$: width in the LWLA implemented with the experimental masses and photon energy. $\Gamma_{Exp}^{PDG}:$ measured widths \cite{PDG18}. $\Gamma_{p/M}^{\left(  Mixed\right)  }:$ width with the experimental photon energy and partially implemented experimental masses. $\Gamma_{p/M}^{\left(The\right)  }$: width without any external input. }
\end{table*}

The similarity of the calculated widths, $\Gamma_{LWLA}^{\left(  The\right)  }$ and $\Gamma_{p/M}^{\left(The\right)  }$, for the considered processes confirms the validity of the LWLA within a few percent of error. On the other hand, their comparison to data, $\Gamma_{Exp}^{PDG},$ third column in the table, makes clear that $\Gamma_{LWLA}^{\left(  The\right)  }$ or $\Gamma_{p/M}^{\left(The\right)  }$ are far from the experimental widths except for$\Upsilon\left(  3S\right)  \rightarrow\gamma\chi_{b2}\left(  2P\right)$. By realizing that this may have to do with the fact that only for this transition the calculated spectral mass difference $M_{3S}-M_{2P}=83$ MeV is very close to the measured one $M_{\Upsilon\left(  3S\right)  }-M_{\chi_{b2}\left(2P\right)  }=86$ MeV, we can try to implement the experimental photon energy $\left\vert \vb*{k}\right\vert _{Exp}$ and the measured mass differences instead of the spectral ones for all the transitions to check whether some improvement can be achieved or not. This implementation can be very easily done in the LWLA since the mass dependence in the amplitude is explicitly factorized. A look at the table shows that the resulting widths that we denote as $\Gamma_{LWLA}^{\left(  The-Exp\right)  }$, second column in the table, are within the error data intervals except for $\Upsilon\left(3S\right)  \rightarrow\gamma\chi_{b0}\left(  2P\right)  $ and $\Upsilon\left(2S\right)  \rightarrow\gamma\chi_{b0}\left(  1P\right)  $ where they are less than a $30\%$ and a $10\%$ off respectively. Keeping always in mind that higher $\frac{\left\vert \vb*{p}_{b}\right\vert }{M_{b}}$ orders might be playing some role, this deviations may indicate some deficiency in the calculated $^{3}\!P_{0}$ wave functions. Actually we could have expected this to occur since our model fits much better the $^{3}\!P_{1,2}$ masses than the $^{3}\!P_{0}$ ones. This means that one should go beyond first order perturbation theory, that gives rise to a mass shift but keeps unaltered the wave function, to get an accurate description of the $^{3}\!P_{0}$ states from our model. This can be confirmed by artificially making the parameters to have slightly different values only for $^{3}\!P_{0}$ states in order to fit their masses. Thus, for instance, taking $\left(  \sigma\right)  _{^{3}\!P_{0}}=875$ MeV fm$^{-1}$ and $\left(  \zeta\right)  _{^{3}\!P_{0}}=120$ MeV fm, the calculated masses and widths are much closer to data.

Hence, we may conclude that the implementation of the experimental masses is an essential ingredient for the explanation of radiative decays.

Following this argumentation we may be confident with our predictions $\Gamma_{LWLA}^{\left(  The-Exp\right)  }$, second column in the table, for $\chi_{b1}\left(  1P\right)  \rightarrow \gamma\Upsilon\left(  1S\right)  $ and $\chi_{b1}\left(  1P\right)\rightarrow\gamma\Upsilon\left(  1S\right)  $ whereas for $\chi_{b0}\left(1P\right)  \rightarrow\gamma\Upsilon\left(  1S\right)  $ we expect a $30\%$ of uncertainty at most. For the sake of comparison let us add that our predicted values are quite in agreement (within a $25\%$ difference) with the ones obtained from other nonrelativistic spectroscopic quark models and from potential nonrelativistic QCD, see Table II in reference \cite{Vai19}. For an alternative theoretical treatment of these decays, see \cite{DeF08}. Regarding $\Upsilon\left(4S\right)  \rightarrow\gamma\chi_{b_{J}}\left(  3P\right)  $ we expect our predicted widths $\Gamma_{LWLA}^{\left(  The-Exp\right)  }$, second column in the table, to be accurate for $\Upsilon\left(4S\right)  \rightarrow\gamma\chi_{b1}\left(  3P\right)  $ and $\Upsilon\left(4S\right)  \rightarrow\gamma\chi_{b2}\left(  3P\right)  $ and more uncertain for $\Upsilon\left(  4S\right)  \rightarrow\gamma\chi_{b0}\left(  3P\right)  $ since the last one is based on an educated guess for the $\chi_{b0}\left(
3P\right)  $ mass.

\bigskip

It is also interesting, for the sake of completeness, to calculate from \eqref{A4}, \eqref{A11} and \eqref{B3} the widths when $\left\vert \vb*{k}\right\vert _{Exp}$ and the experimental masses, when explicitly appearing, are implemented. This means that the measured masses are used in the explicit energy factors entering in the calculation of the width (see \eqref{width} and \eqref{M}) but not in the evaluation of the matrix element in \eqref{M} that still depends implicitly on the calculated spectral masses (this will be detailed in the next section). We call these widths $\Gamma_{p/M}^{\left(  Mixed\right)  }$, fourth column in the table. An inspection of the table makes clear that except for $\Upsilon\left(  3S\right)  \rightarrow\gamma\chi_{b2}\left(  2P\right)  $ where, as explained above, the spectral and experimental mass difference coincides, the widths $\Gamma_{p/M}^{\left(  Mixed\right)  }$ are out of the error data intervals. This points out to the need of making explicit all the mass dependencies in the transition amplitude for their correct experimental implementation if we pretend an accurate decay description beyond the LWLA regime.

\bigskip

Therefore, we have shown that:
\begin{enumerate}[i)]

\item In its range of validity the LWLA, which allows for an easy separation of the mass and wave function dependencies in the transition amplitude, is the more suitable method to give accurate account of the radiative transitions in bottomonium. This is so even for the simplest spectroscopic model, once the experimental masses are properly implemented.

\item The description of radiative decays out of the range of validity of the LWLA requires the explicit factorization of all the mass dependencies in the transition amplitude for its correct experimental implementation.

\end{enumerate}

\section{Beyond the long wave length approximation \label{SV}}

The need of going beyond the LWLA has dealt in the past to the evaluation of some corrections to $\left(  \mathcal{O}_{\alpha}\right)  _{LWLA}$, see for example \cite{Eic05}. Maybe the most common form of the corrected operator is the one where the LWLA mass dependence in the amplitude is preserved while substituting the overlap integral $\int_{0}^{\infty}\dd{r}r^{2}\pqty{R_{n_F L_F}(r)}^*r R_{n_I L_I}(r)$ in \eqref{widthLWLA} by the corrected one (henceforth we shall use $k\equiv\left\vert \vb*{k}\right\vert $).
\begin{multline}
\int_{0}^{\infty}\dd{r}r^{2} \pqty{R_{n_F L_F}(r)}^*\\
\frac{3}{k}\left[  \frac{kr}{2}j_{0}\left(  \frac{kr}{2}\right)  -j_{1}\left(  \frac{kr}{2}\right)  \right]R_{n_I L_I}(r)
\label{ILWLA}
\end{multline}
where $j_{0}$ and $j_{1}$ stand for spherical Bessel functions.

However, this prescription, that reproduces the good description of transitions within the range of applicability of the LWLA when the experimental masses and photon energy are implemented, seems not to work for $\Upsilon\left(  3S\right)  \rightarrow\gamma\chi_{b_J}\left(  1P\right)$, where the LWLA is not valid. In Table~\ref{Tab3s} we show the results from this prescription, $\Gamma_{CLWLA}$, where the subindex CLWLA stands for Corrected Long Wave Length Approximation, for three different nonrelativistic quark models (NRQM), all of them fitting reasonably well the spectrum. Model I is just our model where the experimental masses and $k_{Exp}$ have been used in the calculation of the widths. Model II is another Cornell potential model with a different set of parameter values ($\sigma_{II}=894.66$ MeV/fm, $\zeta_{II}=102.61$ MeV fm and $\left(  M_{b}\right)  _{II}=5180$ MeV apart from a constant fixing the origin of the potential) chosen to get a reasonable fit to the mass centers of gravity of $1S,$ $1P$ and $2S$ states \cite{Eic94}. Model III, see \cite{Seg16} and references therein, contains many more terms in the potential apart from the Cornell ones (spin-spin, spin-orbit, tensor\dots) pretending a unified description of the light and heavy quark meson spectra. (For the sake of completeness we show also results for the measured decays for which the LWLA may be applied.)

\begin{table*}
\centering
\begin{tabular}{ccccc}
\toprule
Radiative Decay & $
\begin{array}{c}
\left(  \Gamma_{CLWLA}\right)  _{I}\\
\text{KeV}
\end{array}
$ & $
\begin{array}
[c]{c}
\left(  \Gamma_{CLWLA}\right)  _{II}\\
\text{KeV}
\end{array}
$ & $
\begin{array}{c}
\left(  \Gamma_{CLWLA}\right)  _{III}\\
\text{KeV}
\end{array}
$ & $
\begin{array}{c}
\Gamma_{Exp}^{PDG}\\
\text{KeV}
\end{array}
$\\
\midrule
$\Upsilon\left(  3S\right)  \rightarrow\gamma\chi_{b_{0}}\left(  1P\right)  $
& $1\times10^{-7}$ & $0.001$ & $0.15$ & $0.054\pm0.013$\\
$\Upsilon\left(  3S\right)  \rightarrow\gamma\chi_{b_{1}}\left(  1P\right)  $
& $0.004$ & $0.008$ & $0.16$ & $0.018\pm0.012$\\
$\Upsilon\left(  3S\right)  \rightarrow\gamma\chi_{b_{2}}\left(  1P\right)  $
& $0.01$ & $0.015$ & $0.08$ & $0.20\pm0.06$\\
\midrule
$\Upsilon\left(  3S\right)  \rightarrow\gamma\chi_{b_{0}}\left(  2P\right)  $
& $1.67$ & $1.35$ & $1.21$ & $1.14\pm0.20$\\
$\Upsilon\left(  3S\right)  \rightarrow\gamma\chi_{b_{1}}\left(  2P\right)  $
& $2.73$ & $2.20$ & $2.13$ & $2.6\pm0.5$\\
$\Upsilon\left(  3S\right)  \rightarrow\gamma\chi_{b_{2}}\left(  2P\right)  $
& $3.02$ & $2.40$ & $2.56$ & $2.7\pm0.5$\\
\midrule
$\Upsilon\left(  2S\right)  \rightarrow\gamma\chi_{b_{0}}\left(  1P\right)  $
& $1.58$ & $1.29$ & $1.09$ & $1.2\pm0.3$\\
$\Upsilon\left(  2S\right)  \rightarrow\gamma\chi_{b_{1}}\left(  1P\right)  $
& $2.43$ & $2.00$ & $1.84$ & $2.2\pm0.3$\\
$\Upsilon\left(  2S\right)  \rightarrow\gamma\chi_{b_{2}}\left(  1P\right)  $
& $2.52$ & $2.04$ & $2.08$ & $2.3\pm0.3$\\
\bottomrule
\end{tabular}
\caption{\label{Tab3s}Calculated widths in the CLWA as compared to data for $\Upsilon\left( 3S\right)  \rightarrow\gamma\chi_{b_J}\left(  1P, 2P\right)  $ and $\Upsilon\left( 2S\right)  \rightarrow\gamma\chi_{b_J}\left(  1P\right)  $. Notation as follows. $\left(  \Gamma_{CLWLA}\right)  _{I}$: width from Model I defined in Sec.~\ref{SII} with the experimental masses and photon energy implemented. $\left(  \Gamma_{CLWLA}\right)  _{II}:$ width from Model II, see \cite{Eic08}. $\left(  \Gamma_{CLWLA}\right)  _{III}$: width from Model III, see \cite{Seg16}. $\Gamma_{Exp}^{PDG}$: measured widths \cite{PDG18}. }
\end{table*}

A glance at the table makes evident that the calculated CLWLA widths are in good agreement with data for processes where the LWLA applies, like $\Upsilon\left(  3S\right)  \rightarrow\gamma\chi_{b_J}\left(  2P\right)$ and  $\Upsilon\left(  2S\right)  \rightarrow\gamma\chi_{b_J}\left(  1P\right)$, but they are in complete disagreement for $\Upsilon\left(  3S\right)  \rightarrow\gamma\chi_{b_J}\left(  1P\right)  $, where the LWLA does not apply. Moreover, in this last case predicted widths for the same decay from different models may differ very much from each other. This points out to an extreme sensitivity of the corrected overlap integral to the details of the wave functions. One could think then of using this sensitivity as a very stringent test of the wave functions. However, before going on with this thought, and according to our discussion in Sec.~\ref{SIV}, one should check whether the assumed mass and wave function dependence separation in the CLWLA should be taken or not for granted. Next we show that it should not and that the difficulties in the description of these decays may be surmounted through a proper factorization of the mass dependencies in the transition amplitude. For this purpose let us consider the matrix element entering in the evaluation of the amplitude \eqref{M} (we may equivalently use $\mathcal{O}_{\alpha}$ or $\mathcal{O}_{\alpha}^{\prime}$).
By denoting
\begin{equation}
\left\vert \Psi\right\rangle \equiv\left\vert J,m,\left(  nL\right)_{b\overline{b}},\left(  S\right)  _{b\overline{b}}\right\rangle
\label{notation}
\end{equation}
we can write the amplitude as
\begin{equation}
\left\langle \mathcal{O}_{\alpha}\right\rangle _{FI}\equiv\left\langle\Psi_{F}\right\vert \mathcal{O}_{\alpha}\left\vert \Psi_{I}\right\rangle=\left\langle \mathcal{O}_{\alpha}\right\rangle _{FI}^{electric}+\left\langle\mathcal{O}_{\alpha}\right\rangle _{FI}^{magnetic}
\label{oalfa}
\end{equation}
where
\begin{equation}
\left\langle \mathcal{O}_{\alpha}\right\rangle _{FI}^{electric}=\left\langle\Psi_{F}\right\vert e^{i(-1)^{\alpha}\left(\frac{\vb*{k}\vdot\vb*{r}}{2}\right)  }\left(  -1\right)^{\alpha}2\vb*{p}\vdot \left(  \vb*{\epsilon}_{\vb*{k}}^{\lambda}\right)  ^{\ast}\left\vert \Psi_{I}\right\rangle
\label{oalfaelec}
\end{equation}
and
\begin{equation}
\left\langle \mathcal{O}_{\alpha}\right\rangle _{FI}^{magnetic}=\left\langle\Psi_{F}\right\vert e^{i(-1)^{\alpha}\left(\frac{\vb*{k}\vdot\vb*{r}}{2}\right)  } i\vb*{\sigma}_{\alpha}\times\vb*{k}\vdot \left(  \vb*{\epsilon}_{\vb*{k}}^{\lambda}\right)  ^{\ast}\left\vert \Psi_{I}\right\rangle
\label{oalfamag}
\end{equation}
In order to extract the mass dependence in $\left\langle \mathcal{O}_{\alpha}\right\rangle _{FI}^{electric}$ we introduce a Parseval identity $\left(\sum_{int}\left\vert \Psi_{int}\right\rangle \left\langle \Psi_{int}\right\vert \right)  $ in terms of eigenstates of the Cornell potential
\begin{multline}
\left\langle \mathcal{O}_{\alpha}\right\rangle _{FI}^{electric}=\sum_{int}\left\langle \Psi_{F}\right\vert e^{i(-1)^{\alpha}\left(\frac{\vb*{k}\vdot\vb*{r}}{2}\right)  }\left\vert \Psi_{int}\right\rangle \\ \left\langle \Psi_{int}\right\vert \left(  -1\right)
^{\alpha}2\vb*{p}\vdot\left(  \vb*{\epsilon}_{\vb*{k}}^{\lambda}\right)  ^{\ast}\left\vert \Psi_{I}\right\rangle
\label{oalfaint}
\end{multline}
Then, substituting $\vb*{p}=-i\frac{M_{b}}{2}\left[\vb*{r},H_{C}\right]  $ we are left with
\begin{multline}
\left\langle \mathcal{O}_{\alpha}\right\rangle _{FI}^{electric}=-iM_{b}\sum_{int}\left\langle \Psi_{F}\right\vert e^{i(-1)^{\alpha
}\left(  \frac{\vb*{k}\vdot\vb*{r}}{2}\right)  }\left\vert\Psi_{int}\right\rangle \\
\left(  M_{I}-M_{int}\right)  \left\langle \Psi_{int}\right\vert (-1)^{\alpha}\vb*
{r}\vdot\left(  \vb*{\epsilon}_{\vb*{k}}^{\lambda}\right)  ^{\ast}\left\vert \Psi_{I}\right\rangle
\label{oalfaelecfinal}
\end{multline}
so that the mass dependencies have been factored out. Notice also that the multiplicative quark mass factor in \eqref{oalfaelecfinal} cancels the same dividing factor in the amplitude \eqref{M}. Therefore, this form of the matrix element preserves the nice feature of separating explicitly the mass and wave function dependencies in the amplitude. Actually, it is trivial to check that for $e^{i\left(-1\right)  ^{\alpha}\left(  \frac{\vb*{k}\vdot\vb*{r}} {2}\right)  }\simeq1$ the LWLA is recovered since then $\left\vert \Psi_{int}\right\rangle =\left\vert \Psi_{I}\right\rangle $ is the only surviving contribution. It should be remarked though that for $e^{i\left(  -1\right)^{\alpha}\left(  \frac{\vb*{k}\vdot\vb*{r}}{2}\right)  }\neq1$ the mass and wave function separation dependence in the amplitude is completely different to the one assumed in the CLWLA. Hence the results obtained from the CLWLA beyond the LWLA regime should not be taken for granted.

It has to be added that for $e^{i(-1)^{\alpha}\left(\frac{\vb*{k}\vdot\vb*{r}}{2}\right)  }\neq1$ there is also a magnetic contribution to the amplitude \eqref{M} which depends on $M_{b}$. Though this introduces an undesired additional model dependence we shall see that for the transitions we are interested in this magnetic contribution has no significant effect on the calculated widths and can be obviated.

\bigskip

For the sake of completeness and convenience for later calculations we also write the resulting expression for $\mathcal{O}_{\alpha}^{\prime}$:
\begin{multline}
\left\langle \mathcal{O}_{\alpha}^{\prime}\right\rangle _{FI}^{electric}= \\
-iM_{b}\sum_{int}\left(  M_{int}-M_{F}\right)  \left\langle \Psi_{F}\right\vert (-1)^{\alpha}\vb*{r}\vdot\left(  \vb*{\epsilon}_{\vb*{k}}^{\lambda}\right)  ^{\ast}\left\vert \Psi_{int}\right\rangle \\
\left\langle \Psi_{int}\right\vert e^{i(-1)^{\alpha}\left(  \frac{\vb*{k}\vdot\vb*{r}}{2}\right)  }\left\vert \Psi_{I}\right\rangle
\label{oprimeelecfinal}
\end{multline}

\bigskip

Expressions \eqref{oalfaelecfinal} and \eqref{oprimeelecfinal} tell us that a good description of a complete set of intermediate states apart form the initial and final ones is needed to accurately reproduce radiative decay widths from a non perfect spectroscopic quark model. Otherwise said, radiative decays are testing the whole spectral model description.

\subsection{${^{3}\!S_{1}}\rightarrow\gamma\,{^{3}\!P_{J}}$ transitions}

Let us apply \eqref{oalfaelecfinal} to the calculation of ${^{3}\!S_{1}}\rightarrow\gamma\,{^{3}\!P_{J}}$ transitions. Working in configuration space and using $\vb*{r}\vdot\left(  \vb*{\epsilon}_{\vb*{k}}^{\lambda}\right)  ^{\ast}=\sqrt{\frac{4\pi}{3}}\left(
Y_{1}^{\lambda}\left(  \widehat{r}\right)  \right)  ^{\ast}r$ and the well known expansion
\begin{equation}
e^{i\left(  -1\right)^{\alpha}\left(  \frac{\vb*{k}\vdot\vb*{r}}{2}\right)  }=\sum_{l=0}^{\infty}\left( i\left(  -1\right)^{\alpha}\right)  ^{l}\sqrt{4\pi}\sqrt{2l+1}j_{l}\left(\frac{kr}{2}\right)  Y_{l}^{0}\left(  \widehat{r}\right)
\end{equation}
the electric part of the amplitude reads, with the same notation as in \eqref{ExpLWLA},
\begin{widetext}
\begin{multline}
\left(  \mathcal{M}_{J_{F},m_{F},J_{I},m_{I}}^{\lambda\text{ }\left(electric\right)  }\right)  ^{{^{3}\!S_{1}}\rightarrow\gamma\,{^{3}\!P_{J}}}=i\sqrt{2M_{I}}\sqrt{2E_{F}}\delta_{S_{I},S_{F}}e_{b}\sum_{l=0}^{\infty}\sum_{n_{int},L_{int},J_{int},m_{int}} (-1)^{l+L_{F}+L_{I}}\left(  4l+1\right)  \left(  2L_{int}+1\right)  \\
  C_{2l,\text{ }J_{int},\text{ }J_{F}}^{0,\text{ }m_{int},\text{ }m_{F}}C_{1,\text{ }J_{int},\text{ }J_{I}}^{\lambda,\text{ }m_{int},\text{ }m_{I}}
\left(\begin{array}{ccc}
L_{int} & 2l & L_{F}\\
0 & 0 & 0
\end{array}\right)
\left(\begin{array}{ccc}
L_{int} & 1 & L_{I}\\
0 & 0 & 0
\end{array}\right) 
\left[\begin{array}{ccc}
J_{F} & 2l & J_{int}\\
L_{int} & S_{F} & L_{F}
\end{array}\right]
\left[\begin{array}{ccc}
J_{I} & 1 & J_{int}\\
L_{int} & S_{I} & L_{I}
\end{array}\right] \\
\left(  M_{I}-M_{int}\right)
\left(  \int_{0}^{\infty}\dd{r}\text{ }r^{2}\left(  R_{n_{F}L_{F}}\right)  ^{\ast}j_{2l}\left(  \frac{kr}{2}\right)  R_{n_{int}L_{int}}\right)
\left(  \int_{0}^{\infty}\dd{r}\text{ }r^{2}\left(  R_{n_{int}L_{int}}\right)  ^{\ast}rR_{n_{I}L_{I}}\right)
\label{FinalSPelectric}
\end{multline}
\end{widetext}

This is our master formula, substituting \eqref{A4}, for a proper implementation of the mass dependencies in the amplitude.

\bigskip

Let us realize that although this formal expression contains a sum over a complete set of intermediate states only a few contributions survive, the ones making the $6j$ symbols to be different from $0$. The underlying reason is that due to the matrix element
\begin{multline}
\left\langle \Psi_{int}\right\vert \vb*{r}\vdot \left(  \vb*{\epsilon}_{\vb*{k}}^{\lambda}\right)  ^{\ast}\left\vert \Psi_{I}\left(  ^{3}\!S_{1}\right)  \right\rangle=\\
\left\langle \Psi_{int}\right\vert \sqrt{\frac{4\pi}{3}}\left(Y_{1}^{\lambda}\left(  \widehat{r}\right)  \right)  ^{\ast}r\left\vert\Psi_{I}\left(  ^{3}\!S_{1}\right)  \right\rangle
\end{multline}
appearing in $\left\langle \mathcal{O}_{\alpha}\right\rangle _{FI}^{electric}$ only intermediate $^{3}\!P_{J_{int}}$ states with $J_{int}=0,1,2$ may give a nonvanishing contribution. Furthermore, from the exponential expansion we see that only the $l=0$ and $l=2$ partial waves contribute to the matrix element $\left\langle \Psi_{F}\left(  ^{3}\!P_{J}\right)  \right\vert e^{i\left(
-1\right)  ^{\alpha}\left(  \frac{\vb*{k}\vdot\vb*{r}}{2}\right)  }\left\vert \Psi_{int}\left(  ^{3}\!P_{J_{int}}\right)\right\rangle$.

\bigskip

From \eqref{FinalSPelectric} for the electric part and \eqref{B3} for the magnetic one the widths are straightforwardly evaluated. In practice, the magnetic contribution hardly plays any role and the sum over intermediate states in the electric part does not need for many terms to converge. More precisely, for $\Upsilon\left(  n_{I}S\right)  \rightarrow\gamma\chi_{b_J}\left(  n_{F}P\right)  $ the consideration of $n_{int}P$ with $n_{int}\leq4$ assures convergence to less than a $2\%$ error when $n_{I}\leq4$ (for the not experimentally measured $^{3}\!P_{0}\left(  3P\right)  $ we have done an educated guess taking it to be $20$ MeV lower than the measured $^{3}\!P_{1}\left(  3P\right)  $ mass; for the not yet measured $^{3}\!P_{0,1,2}\left(  4P\right)  $ resonances we have used the Cornell predicted states from our model; the Cornell wave functions have been used in all cases). For $n_{I}=5$ the same level of convergence requires to include $n_{int}=5$ (for the not yet measured $^{3}\!P_{0,1,2}\left(  5P\right)  $ resonances we have used the Cornell predicted states from our model).

We call the calculated widths $\Gamma_{p/M}^{\left(  The-Exp\right)  }$ consistently with the notation used in Table~\ref{TabwidthsLWLA}. The results from Model I (our model) and Model II are compiled in Table~\ref{TabwidthsSPint}. Notice that \eqref{FinalSPelectric} cannot be consistently applied to Model III since its hamiltonian $H$ contains a spin-orbit term, therefore $\vb*{p}\neq-i\frac{M_{b}}{2}\left[  \vb*{r},H\right] $. Nonetheless, we have checked that for transitions where the calculated mass difference from Model III agrees with data, the results obtained from \eqref{A4} and \eqref{B3} are in good agreement with the ones in Table~\ref{TabwidthsSPint} from Models I and II.

\begin{table*}
\centering
\begin{tabular}{cccc}
\toprule
Radiative Decay & $
\begin{array}{c}
\left(  \Gamma_{p/M}^{\left(  The-Exp\right)  }\right)  _{I}\\
\text{KeV}
\end{array}
$ & $
\begin{array}{c}
\left(  \Gamma_{p/M}^{\left(  The-Exp\right)  }\right)  _{II}\\
\text{KeV}
\end{array}
$ & $
\begin{array}{c}
\Gamma_{Exp}^{PDG}\\
\text{KeV}
\end{array}
$\\
\midrule
$\Upsilon\left(  3S\right)  \rightarrow\gamma\chi_{b_{0}}\left(  1P\right)  $
& $0.08$ & $0.09$ & $0.054\pm0.013$\\
$\Upsilon\left(  3S\right)  \rightarrow\gamma\chi_{b_{1}}\left(  1P\right)  $
& $0.21$ & $0.23$ & $0.018\pm0.012$\\
$\Upsilon\left(  3S\right)  \rightarrow\gamma\chi_{b_{2}}\left(  1P\right)  $
& $0.34$ & $0.34$ & $0.20\pm0.06$\\
\midrule
$\Upsilon\left(  4S\right)  \rightarrow\gamma\chi_{b_{0}}\left(  2P\right)  $
& $0.05$ & $0.05$ & \\
$\Upsilon\left(  4S\right)  \rightarrow\gamma\chi_{b_{1}}\left(  2P\right)  $
& $0.12$ & $0.13$ & \\
$\Upsilon\left(  4S\right)  \rightarrow\gamma\chi_{b_{2}}\left(  2P\right)  $
& $0.20$ & $0.21$ & \\
\midrule
$\Upsilon\left(  4S\right)  \rightarrow\gamma\chi_{b_{0}}\left(  1P\right)  $
& $0.05$ & $0.04$ & \\
$\Upsilon\left(  4S\right)  \rightarrow\gamma\chi_{b_{1}}\left(  1P\right)  $
& $0.11$ & $0.12$ & \\
$\Upsilon\left(  4S\right)  \rightarrow\gamma\chi_{b_{2}}\left(  1P\right)  $
& $0.17$ & $0.17$ & \\
\midrule
$\Upsilon\left(  5S\right)  \rightarrow\gamma\chi_{b_{0}}\left(  3P\right)  $
& $0.08$ & $0.08$  & \\
$\Upsilon\left(  5S\right)  \rightarrow\gamma\chi_{b_{1}}\left(  3P\right)  $
& $0.22$ & $0.22$ & \\
$\Upsilon\left(  5S\right)  \rightarrow\gamma\chi_{b_{2}}\left(  3P\right)  $
& $0.33$ & $0.34$ & \\
\midrule
$\Upsilon\left(  5S\right)  \rightarrow\gamma\chi_{b_{0}}\left(  2P\right)  $
& $0.05$ & $0.07$ & \\
$\Upsilon\left(  5S\right)  \rightarrow\gamma\chi_{b_{1}}\left(  2P\right)  $
& $0.15$ & $0.17$ & \\
$\Upsilon\left(  5S\right)  \rightarrow\gamma\chi_{b_{2}}\left(  2P\right)  $
& $0.21$ & $0.25$ & \\
\midrule
$\Upsilon\left(  5S\right)  \rightarrow\gamma\chi_{b_{0}}\left(  1P\right)  $
& $0.04$ & $0.04$ & \\
$\Upsilon\left(  5S\right)  \rightarrow\gamma\chi_{b_{1}}\left(  1P\right)  $
& $0.09$ & $0.10$ & \\
$\Upsilon\left(  5S\right)  \rightarrow\gamma\chi_{b_{2}}\left(  1P\right)  $
& $0.13$ & $0.14$ &\\
\bottomrule
\end{tabular}
\caption{\label{TabwidthsSPint}Calculated ${^{3}\!S_{1}}\rightarrow\gamma\,{^{3}\!P_{J}}$ widths to order $p/M$ implemented with the experimental masses and photon energy: $\Gamma_{p/M}^{\left(  The-Exp\right)  }$. The widths are evaluated with Models I and II and compared to data when available \cite{PDG18}. Our educated guess for the unknown $\chi_{b0}\left(  3P\right)  $ mass has been $10492$ MeV.}
\end{table*}

As can be checked, the improvement with respect to the CLWLA is enormous. The extreme sensitivity of the results to the wave function used has disappeared and the widths obtained for $\Upsilon\left(  3S\right)  \rightarrow\gamma\chi_{b0}\left(1P\right)  $ and $\Upsilon\left(  3S\right)  \rightarrow\gamma\chi_{b2}\left(1P\right)  $ are much closer to data, being now about a $25\%$ off the experimental intervals. This modest disagreement can be justified in our model from the lack of an accurate wave function description for $^{3}\!P_{0}$ states (notice that they always enter as intermediate states in the calculation of the widths).

As for $\Upsilon\left(  3S\right)  \rightarrow\gamma\chi_{b1}\left(1P\right)  $ our calculated width is one order of magnitude bigger than current data. Moreover, within our Cornell potential model framework the calculated value lies necessarily in between the calculated $\Upsilon\left(3S\right)  \rightarrow\gamma\chi_{b0}\left(  1P\right)  $ and $\Upsilon\left(3S\right)  \rightarrow\gamma\chi_{b2}\left(  1P\right)  $ widths. This is again in contrast with data. Indeed, the experimental situation is rather bizarre as compared to any other $\Upsilon\left(  n_{I}S\right)\rightarrow\gamma\chi_{b_J}\left(  n_{F}P\right)  $ case where the $\Upsilon\left(  n_{I}S\right)  \rightarrow\gamma\chi_{b1}\left(n_{F}P\right)  $ measured width lies always in between those for $\Upsilon\left(  n_{I}S\right)  \rightarrow\gamma\chi_{b0}\left(n_{F}P\right)  $ and $\Upsilon\left(  n_{I}S\right)  \rightarrow\gamma\chi_{b2}\left(  n_{F}P\right)$. Moreover, the experimental relative error in the measurement of the $\Upsilon\left(  3S\right)  \rightarrow\gamma\chi_{b1}\left(  1P\right)  $ width is much larger than for $\Upsilon\left(3S\right)  \rightarrow\gamma\chi_{b0}\left(  1P\right)  $ and $\Upsilon\left(3S\right)  \rightarrow\gamma\chi_{b2}\left(  1P\right)  .$ Then, it would be very important, in our opinion, to refine as much as possible the measurement of the $\Upsilon\left(  3S\right)  \rightarrow\gamma\chi_{b1}\left(2P\right)  $ width to solve this puzzle.

Meantime we think our predictions for not yet measured ${^{3}\!S_{1}}\rightarrow\gamma\,{^{3}\!P_{J}}$ decays, also listed in Table~\ref{TabwidthsSPint}, may be taken as reasonable within a $25\%$ of uncertainty.

\subsection{${^{3}\!P_{J}}\rightarrow\gamma\,{^{3}\!S_{1}}$ transitions}

For ${^{3}\!P_{J}}\rightarrow\gamma\,{^{3}\!S_{1}}$ transitions we proceed in the same manner but using for convenience \eqref{oprimeelecfinal} instead of \eqref{oalfaelecfinal}. The final expression for the electric part of the amplitude element is now

\begin{widetext}
\begin{multline}
\left(  \mathcal{M}_{J_{F},m_{F},J_{I},m_{I}}^{\lambda\text{ }\left(electric\right)  }\right)  ^{{^{3}\!P_{J}}\rightarrow\gamma\,{^{3}\!S_{1}}} =i\sqrt{2M_{I}}\sqrt{2E_{F}}\delta_{S_{I},S_{F}}e_{b}\sum_{l=0}^{\infty}\sum_{n_{int},L_{int},J_{int},m_{int}} (-1)^{l}\left(4l+1\right)  \sqrt{\left(  2L_{I}+1\right)  \left(  2L_{F}+1\right)  } \\
C_{2l,\text{}J_{I},\text{ }J_{int}}^{0,\text{ }m_{I},\text{ }m_{int}}C_{1,\text{ }J_{F},\text{ }J_{int}}^{\lambda,\text{ }m_{F},\text{ }m_{int}}
\left(\begin{array}{ccc}
L_{I} & 2l & L_{int}\\
0 & 0 & 0
\end{array}\right)
\left(\begin{array}{ccc}
L_{F} & 1 & L_{int}\\
0 & 0 & 0
\end{array}\right)
\left[\begin{array}{ccc}
J_{int} & 2l & J_{I}\\
L_{I} & S_{I} & L_{int}
\end{array}\right]
\left[\begin{array}{ccc}
J_{int} & 1 & J_{F}\\
L_{F} & S_{F} & L_{int}
\end{array}\right]\\
\left(  M_{int}-M_{F}\right)
\left(  \int_{0}^{\infty}\dd{r}\text{ }r^{2}\left(  R_{n_{F}L_{F}}\right)  ^{\ast}rR_{n_{int}L_{int}}\right)  \left(  \int_{0}^{\infty}\dd{r}\text{ }r^{2}\left(  R_{n_{int}L_{int}}\right)  ^{\ast}j_{2l}\left(\frac{kr}{2}\right)  R_{n_{I}L_{I}}\right)
\label{FinalPSelectric}
\end{multline}
\end{widetext}
This is our master formula, substituting \eqref{A11}, for a proper implementation of the mass dependencies in the amplitude.

\bigskip

From \eqref{FinalPSelectric} for the electric part and \eqref{B3} for the magnetic we can predict the widths  for not yet measured processes. Our results are shown in Table~\ref{TabwidthsPSint}. Regarding convergence we have used $n_{int}\leq5$ in all cases to assure convergence at the level of $2\%$ error. For the not experimentally measured $^{3}\!P_{0}\left(  3P\right)  $ we have done an educated guess taking it to be $20$ MeV lower than the measured $^{3}\!P_{1}\left(  3P\right)  $ mass; for the not yet measured $^{3}\!P_{0,1,2}\left(4P,5P\right)  $ resonances we have used the Cornell predicted states from our model; the Cornell wave functions have been used in all cases.

\begin{table}
\centering
\begin{tabular}{ccc}
\toprule
Radiative Decay & $
\begin{array}{c}
\left(  \Gamma_{p/M}^{\left(  The-Exp\right)  }\right)  _{I}\\
\text{KeV}
\end{array}
$ & $
\begin{array}{c}
\left(  \Gamma_{p/M}^{\left(  The-Exp\right)  }\right)  _{II}\\
\text{KeV}
\end{array}
$\\
\midrule
$\chi_{b_{0}}\left(  2P\right)  \rightarrow\gamma\Upsilon\left(  2S\right)  $
& $10.08$ & $8.70$\\
$\chi_{b_{1}}\left(  2P\right)  \rightarrow\gamma\Upsilon\left(  2S\right)  $
& $14.06$ & $11.99$\\
$\chi_{b_{2}}\left(  2P\right)  \rightarrow\gamma\Upsilon\left(  2S\right)  $
& $17.07$ & $14.70$\\
\midrule
$\chi_{b_{0}}\left(  2P\right)  \rightarrow\gamma\Upsilon\left(  1S\right)  $
& $9.95$ & $9.08$\\
$\chi_{b_{1}}\left(  2P\right)  \rightarrow\gamma\Upsilon\left(  1S\right)  $
& $11.83$ & $10.77$\\
$\chi_{b_{2}}\left(  2P\right)  \rightarrow\gamma\Upsilon\left(  1S\right)  $
& $14.76$ & $13.22$\\
\midrule
$\chi_{b_{0}}\left(  3P\right)  \rightarrow\gamma\Upsilon\left(  3S\right)  $
& $5.23$ & $4.81$\\
$\chi_{b_{1}}\left(  3P\right)  \rightarrow\gamma\Upsilon\left(  3S\right)  $
& $8.33$ & $7.25$\\
$\chi_{b_{2}}\left(  3P\right)  \rightarrow\gamma\Upsilon\left(  3S\right)  $
& $10.63$ & $9.24$\\
\midrule
$\chi_{b_{0}}\left(  3P\right)  \rightarrow\gamma\Upsilon\left(  2S\right)  $
& $3.99$ & $3.69$\\
$\chi_{b_{1}}\left(  3P\right)  \rightarrow\gamma\Upsilon\left(  2S\right)  $
& $4.82$ & $4.42$\\
$\chi_{b_{2}}\left(  3P\right)  \rightarrow\gamma\Upsilon\left(  2S\right)  $
& $5.80$ & $5.27$\\
\midrule
$\chi_{b_{0}}\left(  3P\right)  \rightarrow\gamma\Upsilon\left(  1S\right)  $
& $5.51$ & $5.31$\\
$\chi_{b_{1}}\left(  3P\right)  \rightarrow\gamma\Upsilon\left(  1S\right)  $
& $6.57$ & $6.25$\\
$\chi_{b_{2}}\left(  3P\right)  \rightarrow\gamma\Upsilon\left(  1S\right)  $
& $8.45$ & $7.88$\\
\bottomrule
\end{tabular}
\caption{\label{TabwidthsPSint}Calculated ${^{3}\!P_{J}}\rightarrow\gamma\,{^{3}\!S_{1}}$ widths to order $p/M$ implemented with the experimental masses and photon energy: $\Gamma_{p/M}^{\left(  The-Exp\right)  }$. Our educated guess for the unknown $\chi_{b0}\left(  3P\right)  $ mass has been $10492$ MeV.}
\end{table}
(Notice that one could alternatively choose \eqref{oalfaelecfinal} for ${^{3}\!P_{J}}\rightarrow\gamma\,{^{3}\!S_{1}}$ transitions (or \eqref{oprimeelecfinal} for ${^{3}\!S_{1}}\rightarrow\gamma\,{^{3}\!P_{J}}$ ones). The only difference is in the set of intermediate contributing states that would be formed by $S$ and $D$ waves. This adds support to our former assertion that radiative decays may serve as a stringent test of the whole spectral model description.)

These predictions and the ones in Table~\ref{TabwidthsSPint} are the main results of our research. Their comparison to future data will be a definite test of the proposed formalism to deal with radiative decays beyond the LWLA.

\section{Summary\label{SVI}}

Starting from a simple nonrelativistic quark potential model fitting well the low lying spin triplet $1^{--}$ and $2^{++}$ (and to a lesser extent $1^{++}$) bottomonium spectroscopy we have calculated ${^{3}\!S_{1}}\rightarrow\gamma\,{^{3}\!P_{J}}$ and ${^{3}\!P_{J}}\rightarrow\gamma\,{^{3}\!S_{1}}$ decay widths by using a nonrelativistic reduction, up to $\frac{\left\vert \vb*{p}_{b}\right\vert }{M_{b}}$ order, of the Elementary Emission Model transition operator. In this decay model the emission of the photon is assumed to take place by the quark or the antiquark of the decaying meson. A great simplification applies when the wave length of the emitted photon is much larger than the hadronic size scale for the transition. This occurs for example for decays involving the lowest lying spectral states. Then, in this Long Wave Length Approximation (LWLA) the amplitude dependence on the mass and wave function of the initial and final mesons can be factored out. This permits a step by step analysis of the requirements needed to get an accurate description of data from a spectroscopic potential model. As a general result, we have shown that the implementation of the experimental masses and photon energy, instead of the calculated ones, in the evaluation of the transition amplitude is an essential requirement for predictions to be in accord with data. This implementation is justified under the assumption that the difference between the measured masses and the calculated ones corresponds in most cases to a first order perturbative effect. The comparison of the resulting widths with data support this assumption since the only modest $\left(  25\%\right)  $ deviation from data corresponds to transitions involving $0^{++}$ states for which the difference between the calculated and measured masses is significantly bigger than for the $2^{++}$ and $1^{++}$ cases.

\bigskip

For general transitions between bottomonium states where the LWLA does not necessarily apply a new method to factor out the mass and wave function dependence of the amplitude has been developed and applied to $^{3}\!S_{1}\longleftrightarrow{^{3}\!P_{J}}$ transitions. This method is based on the introduction of a complete set of intermediate Cornell states in the calculation of the amplitude. Thus, for instance, the ${^{3}\!S_{1}}\rightarrow\gamma\,{^{3}\!P_{J}}$ amplitude can be written as a sum of LWLA like amplitudes from the initial to intermediate $P-$ wave states with coefficients depending on the intermediate and final states. The introduction of intermediate states for an accurate description of the decay widths from a non perfect spectroscopic model indicates that radiative decays beyond the LWLA may serve as a very stringent test of the spectroscopic wave functions. As a matter of fact, any inaccuracy in the calculation of ${^{3}\!S_{1}}\rightarrow\gamma\,{^{3}\!P_{J}}$ amplitudes for which the LWLA applies translates into an inaccuracy for general ${^{3}\!S_{1}}\rightarrow\gamma\,{^{3}\!P_{J}}$ transitions beyond the LWLA. From the scarce data available we have verified that the same level of inaccuracy $\left(  25\%\right)  $ may be expected in both cases. This makes us confident in our predictions for not yet measured decay widths which may serve as a guide for future experimental searches.

\bigskip

In summary, we have developed a formalism to get an accurate description of the electromagnetic $^{3}\!S_{1}\longleftrightarrow{^{3}\!P_{J}}$ bottomonium transition widths from a $\frac{\left\vert \vb*{p}_{b}\right\vert}{M_{b}}$ order Elementary Emission Decay Model and a simple nonrelativistic spectroscopic Cornell potential model. Our formalism can be used for more refined nonrelativistic potentials as far as they only depend on the quark-antiquark separation. Couple channel corrections whose effect has been partially analyzed in unquenched quark models \cite{Fer13,*Fer14_5,*Fer14_9} can be also incorporated trough the correction to the wavefunctions. However, the formalism cannot be easily generalized to charmonium where it can be checked that higher orders in $\frac{\left\vert\vb*{p}_{c}\right\vert }{M_{c}}$ play an important role. Work along this line is in progress.

\begin{acknowledgments}
This work has been supported by \foreignlanguage{spanish}{Ministerio de Economía y Competitividad} of Spain (MINECO) and EU Feder grant FPA2016-77177-C2-1-P and by SEV-2014-0398. R.\,B.\ acknowledges a FPI fellowship from the \foreignlanguage{spanish}{Ministerio de Ciencia, Innovación y Universidades} of Spain under grant BES-2017-079860. We are grateful to J.\,Segovia for providing us with a set of wavefuctions for comparison.
\end{acknowledgments}

\appendix

\section{Electric transitions \label{SVII}}

Electric transitions are driven by the $\vb*{p}-$ dependent term in the transition operator \eqref{O}. For the ${^{3}\!S_{1}}\rightarrow\gamma\,{^{3}\!P_{J}}$ case we use for convenience the electric part of \eqref{Ored}
\begin{equation}
\left(  \mathcal{O}_{\alpha}\right)  ^{electric}= e^{i\left(  -1\right)^{\alpha}\left(  \frac{\vb*{k}\vdot\vb*{r}}{2}\right)
}(-1)^{\alpha}2\vb*{p}\vdot\left(  \vb*{\epsilon}_{\vb*{k}}^{\lambda}\right)^{\ast}
\label{A1}
\end{equation}
so that the amplitude can be written as
\begin{multline}
\mathcal{M}_{J_{F},m_{F},J_{I},m_{I}}^{\lambda\text{ }\left(  electric\right)}=\sqrt{2M_{I}}\sqrt{2E_{F}}\sum_{\alpha=1,2}\frac{e_{\alpha}}{2M_{b}}\\
\left\langle J_{F},m_{F},\left(  n_{F}L_{F}\right)  _{b\overline{b}},\left(S_{F}\right)  _{b\overline{b}}\right\vert \\
\left(  \mathcal{O}_{\alpha}\right)^{electric}\left\vert J_{I},m_{I},\left(  n_{I}L_{I}\right)  _{b\overline{b}},\left(  S_{I}\right)  _{b\overline{b}}\right\rangle
\label{A2}
\end{multline}

In configuration space $\vb*{p}\hookrightarrow-i\vb*{\nabla}.$ As the initial state, $L_{I}=0,$ has no angular dependence, and the photon travels along the Z axis one has
\begin{equation}
\vb*{p}\vdot\left(  \vb*{\epsilon}_{\vb*{k}}^{\lambda}\right)^{\ast}\left\vert \left(  n_{I}L_{I}\right)  _{b\overline{b}}\right\rangle \hookrightarrow-i\sqrt{\frac{4\pi}{3}}\pqty{Y_{1}^{\lambda}\left(\widehat{r}\right)}^*  \frac{dR_{n_{I}L_{I}}}{\dd{r}}\frac{1}{\sqrt{4\pi}}
\label{A3}
\end{equation}
where $R_{n_{I}L_{I}}$ is the radial wave function of the initial state. Then, using the expansion of the exponential

\begin{equation}
e^{i\left(  -1\right)^{\alpha}\left(  \frac{\vb*{k}\vdot\vb*{r}}{2}\right)  }=\sum_{l=0}^{\infty}\left( i\left(  -1\right)^{\alpha}\right)  ^{l}\sqrt{4\pi}\sqrt{2l+1}j_{l}\left(\frac{kr}{2}\right)  Y_{l}^{0}\left(  \widehat{r}\right)
\end{equation}
and some angular momentum algebra one gets
\begin{multline}
\left(  \mathcal{M}_{J_{F},m_{F},J_{I},m_{I}}^{\lambda\text{ }\left(electric\right)  }\right)  ^{{^{3}\!S_{1}}\rightarrow\gamma\,{^{3}\!P_{J}}}  =\\
\sqrt{2M_{I}}\sqrt{2E_{F}}\delta_{S_{I},S_{F}}\frac{e_{b}}{M_{b}}\sum_{l=0}^{\infty}\left(  1-(-1)^{l}\right) \\
 \left(  \mathcal{I}_{l-1}\left(  \frac{k}{2}\right)  +\mathcal{I}_{l+1}\left(  \frac{k}{2}\right)  \right)\\
 i^{l+1}B_{l,L_{F}}C_{l,\;J_{F},\;J_{I}}^{\lambda,m_{F},m_{I}}
\left(\begin{array}{ccc}
L_{F} & l & L_{I}\\
0 & 0 & 0
\end{array}\right)
\left[\begin{array}{ccc}
J_{I} & l & J_{F}\\
L_{F} & S_{F} & L_{I}
\end{array}\right]
\label{A4}
\end{multline}
where
\begin{multline}
\mathcal{I}_{l\mp1}\left(  \frac{k}{2}\right)  \equiv\int_{0}^{\infty}\dd{r}\text{ }r^{2}\\\left(  R_{n_{F}L_{F}}\right)  ^{\ast}\text{ }j_{l\mp1}\left(\frac{kr}{2}\right)  \pqty{-i\dv{R_{n_{I}L_{I}}}{r}}
\label{A5}
\end{multline}
\begin{align}
B_{l,L_{F}}&\equiv(-1)^{L_{F}+1}\sqrt{\frac{l\left(  l+1\right)\left(  2L_{F}+1\right)  }{2}} \label{A6}\\
C_{l,\;J_{F},\;J_{I}}^{\lambda,m_{F},m_{I}}&\equiv\left(-1\right)  ^{J_{F}-l-m_{I}}\sqrt{2J_{I}+1}
\left(\begin{array}{ccc}
l & J_{F} & J_{I}\\
\lambda & m_{F} & m_{I}
\end{array}\right)
\label{A7}
\end{align}
and
\begin{multline}
\left[\begin{array}{ccc}
j_1 & j_2 & j_{12}\\
j_3 & j & j_{23}
\end{array}\right]
\equiv(-1)^{j_1+j_2+j_3+j}\\
\sqrt{\left(2j_{12}+1\right)  \left(  2 j_{23}+1\right)  }
\left\{\begin{array}{ccc}
j_1 & j_2 & j_{12}\\
j_3 & j & j_{23}
\end{array}\right\}
\label{A8}
\end{multline}
with $\left\{  {}\right\}  $ standing for the $6j$ symbol.

\bigskip

As for the ${^{3}\!P_{J}}\rightarrow\gamma\,{^{3}\!S_{1}}$ case we use for convenience
\begin{equation}
\left(  \mathcal{O}_{\alpha}^{\prime}\right)  ^{electric}= (-1)^{\alpha}2\vb*{p}\vdot\left(\vb*{\epsilon}_{\vb*{k}}^{\lambda}\right)  ^{\ast}e^{i(-1)^{\alpha}\left(\frac{\vb*{k}\vdot\vb*{r}}{2}\right)  } \label{A9}
\end{equation}
so that the amplitude can be written as
\begin{multline}
\mathcal{M}_{J_{F},m_{F},J_{I},m_{I}}^{\lambda\text{ }\left(  electric\right)}=\sqrt{2M_{I}}\sqrt{2E_{F}}\sum_{\alpha=1,2}\frac{e_{\alpha}}{2M_{b}} \\
\left\langle J_{F},m_{F},\left(  n_{F}L_{F}\right)  _{b\overline{b}},\left(S_{F}\right)  _{b\overline{b}}\right\vert \\
\left(  \mathcal{O}_{\alpha}^{\prime}\right)  ^{electric}\left\vert J_{I},m_{I},\left(  n_{I}L_{I}\right)_{b\overline{b}},\left(  S_{I}\right)  _{b\overline{b}}\right\rangle
\label{A10}
\end{multline}
By proceeding as above one gets
\begin{multline}
\left(  \mathcal{M}_{J_{F},m_{F},J_{I},m_{I}}^{\lambda\text{ }\left(electric\right)  }\right)  ^{{^{3}\!P_{J}}\rightarrow\gamma\,{^{3}\!S_{1}}} 
=\\
\sqrt{2M_{I}}\sqrt{2E_{F}}\delta_{S_{I},S_{F}}\frac{e_{b}}{M_{b}}\sum_{l=0}^{\infty}\left(  1-(-1)^{l}\right) \\
\left(  \mathcal{J}_{l-1}\left(  \frac{k}{2}\right)  +\mathcal{J}_{l+1}\left(  \frac{k}{2}\right)  \right) \\
i^{l+1}B_{l,L_{F}}C_{l,\;J_{F},\;J_{I}}^{\lambda,m_{F},m_{I}}
\left(\begin{array}{ccc}
L_{F} & l & L_{I}\\
0 & 0 & 0
\end{array}\right)
\left[\begin{array}{ccc}
J_{I} & l & J_{F}\\
L_{F} & S_{F} & L_{I}
\end{array}\right]
\label{A11}
\end{multline}
where
\begin{multline}
\mathcal{J}_{l\mp1}\left(  \frac{k}{2}\right)  \equiv\int_{0}^{\infty}\dd{r}\text{ }r^{2}\\
\left(  -i\dv{R_{n_{F}L_{F}}}{r}\right)  ^{\ast}\text{ }j_{l\mp1}\left(  \frac{kr}{2}\right)  \text{ }R_{n_{I}L_{I}}
\label{A12}
\end{multline}

\section{Magnetic transitions \label{SVIII}}

The magnetic transitions are driven by the $\vb*{\sigma}-$ dependent term in the transition operator \eqref{O}
\begin{equation}
\left(  \mathcal{O}_{\alpha}\right)  ^{magnetic}=i\vb*{\sigma}_{\alpha}\times\vb*{k}\vdot\left(  \vb*{\epsilon}_{\vb*{k}}^{\lambda}\right)  ^{\ast}e^{i(-1)^{\alpha}\left(  \frac{\vb*{k}\vdot\vb*{r}}{2}\right)  }
\label{B1}
\end{equation}
so that the amplitude can be written as
\begin{multline}
\mathcal{M}_{J_{F},m_{F},J_{I},m_{I}}^{\lambda\text{ }\left(  magnetic\right)}=\sqrt{2M_{I}}\sqrt{2E_{F}}\sum_{\alpha=1,2}\frac{e_{\alpha}}{2M_{b}}\\
\left\langle J_{F},m_{F},\left(  n_{F}L_{F}\right)  _{b\overline{b}},\left(S_{F}\right)  _{b\overline{b}}\right\vert \\
\left(  \mathcal{O}_{\alpha}\right)^{magnetic}\left\vert J_{I},m_{I},\left(  n_{I}L_{I}\right)  _{b\overline{b}},\left(  S_{I}\right)  _{b\overline{b}}\right\rangle
\label{B2}
\end{multline}
A straightforward but lengthy calculation yields
\begin{multline}
\mathcal{M}_{J_{F},m_{F},J_{I},m_{I}}^{\lambda\text{ }\left(  magnetic\right)} =\sqrt{2M_{I}}\sqrt{2E_{F}}\frac{e_{b}}{M_{b}}\lambda k \\
\sum_{l=1}^{\infty}\left(  (-1)^{l+1}+\left(  -1\right)^{S_{F}-S_{I}+1}\right) \\
\mathcal{K}_{l-1}\left(  \frac{k}{2}\right)i^{l+1} D_{l,L_{F}} \\
\left(\begin{array}{ccc}
L_{F} & l-1 & L_{I}\\
0 & 0 & 0
\end{array}\right)
\left[\begin{array}{ccc}
S_{I} & 1 & S_{F}\\
\frac{1}{2} & \frac{1}{2} & \frac{1}{2}
\end{array}\right]
A
\label{B3}
\end{multline}
where
\begin{equation}
\mathcal{K}_{l-1}\left(  \frac{k}{2}\right)  \equiv\int_{0}^{\infty}\dd{r}\text{ }r^{2}\left(  R_{n_{F}L_{F}}\right)  ^{\ast}\text{ }j_{l-1}\left(\frac{kr}{2}\right)  \text{ }R_{n_{I}L_{I}}
\label{B4}
\end{equation}
\begin{equation}
D_{l,L_{F}}\equiv\frac{(-1)^{L_{F}+l+1}}{2}\sqrt{\frac{3\left(2l-1\right)  \left(  \left(  2L_{F}+1\right)  \right)  }{2}}
\label{B5}
\end{equation}

\begin{multline}
A  \equiv\sqrt{l+1}C_{l,\;J_{F},\;J_{I}}^{\lambda,m_{F},m_{I}}
\left[\begin{array}{ccc}
L_{F} & l-1 & L_{I}\\
S_{F} & 1 & S_{I}\\
J_{F} & l & J_{I}
\end{array}\right]  \\
-\lambda\sqrt{2l-1}C_{l-1,J_{F},\;J_{I}}^{\lambda,\;m_{F},m_{I}}
\left[\begin{array}{ccc}
L_{F} & l-1 & L_{I}\\
S_{F} & 1 & S_{I}\\
J_{F} & l-1 & J_{I}
\end{array}\right] \\
+\sqrt{l-2}C_{l-2,J_{F},\;J_{I}}^{\lambda,\;m_{F},m_{I}}
\left[\begin{array}{ccc}
L_{F} & l-1 & L_{I}\\
S_{F} & 1 & S_{I}\\
J_{F} & l-2 & J_{I}
\end{array}\right]
\label{B6}
\end{multline}
and
\begin{multline}
\left[\begin{array}{ccc}
j_{1} & j_{2} & j_{12}\\
j_{3} & j_{4} & j_{34}\\
j_{13} & j_{24} & j
\end{array}\right]
\equiv\\
\sqrt{\left(  2 j_{12}+1\right)  \left(  2 j_{34}+1\right)  \left(2 j_{13}+1\right)  \left(  2 j_{24}+1\right)  } \\
\left\{\begin{array}{ccc}
j_{1} & j_{2} & j_{12}\\
j_{3} & j_{4} & j_{34}\\
j_{13} & j_{24} & j
\end{array}\right\}
\label{B7}
\end{multline}
with $\left\{  {}\right\}  $ standing for the $9j$ symbol.

\bibliography{embottbib}

\begin{thebibliography}{20}%
\makeatletter
\providecommand \@ifxundefined [1]{%
 \@ifx{#1\undefined}
}%
\providecommand \@ifnum [1]{%
 \ifnum #1\expandafter \@firstoftwo
 \else \expandafter \@secondoftwo
 \fi
}%
\providecommand \@ifx [1]{%
 \ifx #1\expandafter \@firstoftwo
 \else \expandafter \@secondoftwo
 \fi
}%
\providecommand \natexlab [1]{#1}%
\providecommand \enquote  [1]{``#1''}%
\providecommand \bibnamefont  [1]{#1}%
\providecommand \bibfnamefont [1]{#1}%
\providecommand \citenamefont [1]{#1}%
\providecommand \href@noop [0]{\@secondoftwo}%
\providecommand \href [0]{\begingroup \@sanitize@url \@href}%
\providecommand \@href[1]{\@@startlink{#1}\@@href}%
\providecommand \@@href[1]{\endgroup#1\@@endlink}%
\providecommand \@sanitize@url [0]{\catcode `\\12\catcode `\$12\catcode
  `\&12\catcode `\#12\catcode `\^12\catcode `\_12\catcode `\%12\relax}%
\providecommand \@@startlink[1]{}%
\providecommand \@@endlink[0]{}%
\providecommand \url  [0]{\begingroup\@sanitize@url \@url }%
\providecommand \@url [1]{\endgroup\@href {#1}{\urlprefix }}%
\providecommand \urlprefix  [0]{URL }%
\providecommand \Eprint [0]{\href }%
\providecommand \doibase [0]{http://dx.doi.org/}%
\providecommand \selectlanguage [0]{\@gobble}%
\providecommand \bibinfo  [0]{\@secondoftwo}%
\providecommand \bibfield  [0]{\@secondoftwo}%
\providecommand \translation [1]{[#1]}%
\providecommand \BibitemOpen [0]{}%
\providecommand \bibitemStop [0]{}%
\providecommand \bibitemNoStop [0]{.\EOS\space}%
\providecommand \EOS [0]{\spacefactor3000\relax}%
\providecommand \BibitemShut  [1]{\csname bibitem#1\endcsname}%
\let\auto@bib@innerbib\@empty
\bibitem [{\citenamefont {Le~Yaouanc}\ \emph {et~al.}(1988)\citenamefont
  {Le~Yaouanc}, \citenamefont {Oliver}, \citenamefont {Pene},\ and\
  \citenamefont {Raynal}}]{LeY88}%
  \BibitemOpen
  \bibfield  {author} {\bibinfo {author} {\bibfnamefont {A.}~\bibnamefont
  {Le~Yaouanc}}, \bibinfo {author} {\bibfnamefont {L.}~\bibnamefont {Oliver}},
  \bibinfo {author} {\bibfnamefont {O.}~\bibnamefont {Pene}}, \ and\ \bibinfo
  {author} {\bibfnamefont {J.~C.}\ \bibnamefont {Raynal}},\ }\href@noop {}
  {\emph {\bibinfo {title} {{Hadron transitions in the quark model}}}}\
  (\bibinfo  {publisher} {New York, USA: Gordon and Breach},\ \bibinfo {year}
  {1988})\BibitemShut {NoStop}%
\bibitem [{\citenamefont {Brambilla}\ \emph {et~al.}(2004)\citenamefont
  {Brambilla} \emph {et~al.}}]{Eic05}%
  \BibitemOpen
  \bibfield  {author} {\bibinfo {author} {\bibfnamefont {N.}~\bibnamefont
  {Brambilla}} \emph {et~al.} (\bibinfo {collaboration} {Quarkonium Working
  Group}),\ }\href@noop {} {\enquote {\bibinfo {title} {{Heavy quarkonium
  physics}},}\ } (\bibinfo {year} {2004}),\ \Eprint
  {http://arxiv.org/abs/hep-ph/0412158} {arXiv:hep-ph/0412158 [hep-ph]}
  \BibitemShut {NoStop}%
\bibitem [{\citenamefont {Godfrey}\ and\ \citenamefont {Isgur}(1985)}]{GI85}%
  \BibitemOpen
  \bibfield  {author} {\bibinfo {author} {\bibfnamefont {S.}~\bibnamefont
  {Godfrey}}\ and\ \bibinfo {author} {\bibfnamefont {N.}~\bibnamefont
  {Isgur}},\ }\href {\doibase 10.1103/PhysRevD.32.189} {\bibfield  {journal}
  {\bibinfo  {journal} {Phys. Rev.D}\ }\textbf {\bibinfo {volume} {32}},\
  \bibinfo {pages} {189} (\bibinfo {year} {1985})}\BibitemShut {NoStop}%
\bibitem [{\citenamefont {Segovia}\ \emph {et~al.}(2016)\citenamefont
  {Segovia}, \citenamefont {Ortega}, \citenamefont {Entem},\ and\ \citenamefont
  {Fernández}}]{Seg16}%
  \BibitemOpen
  \bibfield  {author} {\bibinfo {author} {\bibfnamefont {J.}~\bibnamefont
  {Segovia}}, \bibinfo {author} {\bibfnamefont {P.~G.}\ \bibnamefont {Ortega}},
  \bibinfo {author} {\bibfnamefont {D.~R.}\ \bibnamefont {Entem}}, \ and\
  \bibinfo {author} {\bibfnamefont {F.}~\bibnamefont {Fernández}},\ }\href
  {\doibase 10.1103/PhysRevD.93.074027} {\bibfield  {journal} {\bibinfo
  {journal} {Phys. Rev.D}\ }\textbf {\bibinfo {volume} {93}},\ \bibinfo {pages}
  {074027} (\bibinfo {year} {2016})},\ \Eprint
  {http://arxiv.org/abs/1601.05093} {arXiv:1601.05093 [hep-ph]} \BibitemShut
  {NoStop}%
\bibitem [{\citenamefont {Eichten}\ \emph {et~al.}(1978)\citenamefont
  {Eichten}, \citenamefont {Gottfried}, \citenamefont {Kinoshita},
  \citenamefont {Lane},\ and\ \citenamefont {Yan}}]{Eic80}%
  \BibitemOpen
  \bibfield  {author} {\bibinfo {author} {\bibfnamefont {E.}~\bibnamefont
  {Eichten}}, \bibinfo {author} {\bibfnamefont {K.}~\bibnamefont {Gottfried}},
  \bibinfo {author} {\bibfnamefont {T.}~\bibnamefont {Kinoshita}}, \bibinfo
  {author} {\bibfnamefont {K.~D.}\ \bibnamefont {Lane}}, \ and\ \bibinfo
  {author} {\bibfnamefont {T.-M.}\ \bibnamefont {Yan}},\ }\href {\doibase
  10.1103/PhysRevD.17.3090} {\bibfield  {journal} {\bibinfo  {journal} {Phys.
  Rev.D}\ }\textbf {\bibinfo {volume} {17}},\ \bibinfo {pages} {3090} (\bibinfo
  {year} {1978})},\ \bibinfo {note} {[Erratum: Phys.
  Rev.D21,313(1980)]}\BibitemShut {NoStop}%
\bibitem [{\citenamefont {Eichten}\ and\ \citenamefont {Quigg}(1994)}]{Eic94}%
  \BibitemOpen
  \bibfield  {author} {\bibinfo {author} {\bibfnamefont {E.~J.}\ \bibnamefont
  {Eichten}}\ and\ \bibinfo {author} {\bibfnamefont {C.}~\bibnamefont
  {Quigg}},\ }\href {\doibase 10.1103/PhysRevD.49.5845} {\bibfield  {journal}
  {\bibinfo  {journal} {Phys. Rev.D}\ }\textbf {\bibinfo {volume} {49}},\
  \bibinfo {pages} {5845} (\bibinfo {year} {1994})},\ \Eprint
  {http://arxiv.org/abs/hep-ph/9402210} {arXiv:hep-ph/9402210 [hep-ph]}
  \BibitemShut {NoStop}%
\bibitem [{\citenamefont {Bali}(2001)}]{Bal01}%
  \BibitemOpen
  \bibfield  {author} {\bibinfo {author} {\bibfnamefont {G.~S.}\ \bibnamefont
  {Bali}},\ }\href {\doibase 10.1016/S0370-1573(00)00079-X} {\bibfield
  {journal} {\bibinfo  {journal} {Phys. Rep.}\ }\textbf {\bibinfo {volume}
  {343}},\ \bibinfo {pages} {1} (\bibinfo {year} {2001})},\ \Eprint
  {http://arxiv.org/abs/hep-ph/0001312} {arXiv:hep-ph/0001312 [hep-ph]}
  \BibitemShut {NoStop}%
\bibitem [{\citenamefont {Titard}\ and\ \citenamefont
  {Yndurain}(1995)}]{Ynd95}%
  \BibitemOpen
  \bibfield  {author} {\bibinfo {author} {\bibfnamefont {S.}~\bibnamefont
  {Titard}}\ and\ \bibinfo {author} {\bibfnamefont {F.~J.}\ \bibnamefont
  {Yndurain}},\ }\href {\doibase 10.1016/0370-2693(95)00429-O} {\bibfield
  {journal} {\bibinfo  {journal} {Phys. Lett.B}\ }\textbf {\bibinfo {volume}
  {351}},\ \bibinfo {pages} {541} (\bibinfo {year} {1995})},\ \Eprint
  {http://arxiv.org/abs/hep-ph/9501338} {arXiv:hep-ph/9501338 [hep-ph]}
  \BibitemShut {NoStop}%
\bibitem [{\citenamefont {Gonzalez}(2014)}]{Gon14}%
  \BibitemOpen
  \bibfield  {author} {\bibinfo {author} {\bibfnamefont {P.}~\bibnamefont
  {Gonzalez}},\ }\href {\doibase 10.1088/0954-3899/41/9/095001} {\bibfield
  {journal} {\bibinfo  {journal} {J. Phys.G}\ }\textbf {\bibinfo {volume}
  {41}},\ \bibinfo {pages} {095001} (\bibinfo {year} {2014})},\ \Eprint
  {http://arxiv.org/abs/1406.5025} {arXiv:1406.5025 [hep-ph]} \BibitemShut
  {NoStop}%
\bibitem [{\citenamefont {Tanabashi}\ \emph {et~al.}(2018)\citenamefont
  {Tanabashi} \emph {et~al.}}]{PDG18}%
  \BibitemOpen
  \bibfield  {author} {\bibinfo {author} {\bibfnamefont {M.}~\bibnamefont
  {Tanabashi}} \emph {et~al.} (\bibinfo {collaboration} {Particle Data
  Group}),\ }\href {\doibase 10.1103/PhysRevD.98.030001} {\bibfield  {journal}
  {\bibinfo  {journal} {Phys. Rev.D}\ }\textbf {\bibinfo {volume} {98}},\
  \bibinfo {pages} {030001} (\bibinfo {year} {2018})},\ \bibinfo {note} {[And
  2019 update]}\BibitemShut {NoStop}%
\bibitem [{\citenamefont {Bruschini}\ and\ \citenamefont
  {González}(2019{\natexlab{a}})}]{Bru19}%
  \BibitemOpen
  \bibfield  {author} {\bibinfo {author} {\bibfnamefont {R.}~\bibnamefont
  {Bruschini}}\ and\ \bibinfo {author} {\bibfnamefont {P.}~\bibnamefont
  {González}},\ }\href {\doibase 10.1016/j.physletb.2019.03.017} {\bibfield
  {journal} {\bibinfo  {journal} {Phys. Lett.B}\ }\textbf {\bibinfo {volume}
  {791}},\ \bibinfo {pages} {409} (\bibinfo {year} {2019}{\natexlab{a}})},\
  \Eprint {http://arxiv.org/abs/1811.08236} {arXiv:1811.08236 [hep-ph]}
  \BibitemShut {NoStop}%
\bibitem [{\citenamefont {González}(2015)}]{Gon15}%
  \BibitemOpen
  \bibfield  {author} {\bibinfo {author} {\bibfnamefont {P.}~\bibnamefont
  {González}},\ }\href {\doibase 10.1103/PhysRevD.92.014017} {\bibfield
  {journal} {\bibinfo  {journal} {Phys. Rev.D}\ }\textbf {\bibinfo {volume}
  {92}},\ \bibinfo {pages} {014017} (\bibinfo {year} {2015})},\ \Eprint
  {http://arxiv.org/abs/1507.02397} {arXiv:1507.02397 [hep-ph]} \BibitemShut
  {NoStop}%
\bibitem [{\citenamefont {Bruschini}\ and\ \citenamefont
  {González}(2019{\natexlab{b}})}]{BrG19}%
  \BibitemOpen
  \bibfield  {author} {\bibinfo {author} {\bibfnamefont {R.}~\bibnamefont
  {Bruschini}}\ and\ \bibinfo {author} {\bibfnamefont {P.}~\bibnamefont
  {González}},\ }\href {\doibase 10.1103/PhysRevC.99.045205} {\bibfield
  {journal} {\bibinfo  {journal} {Phys. Rev.C}\ }\textbf {\bibinfo {volume}
  {99}},\ \bibinfo {pages} {045205} (\bibinfo {year} {2019}{\natexlab{b}})},\
  \Eprint {http://arxiv.org/abs/1904.02978} {arXiv:1904.02978 [hep-ph]}
  \BibitemShut {NoStop}%
\bibitem [{\citenamefont {Eichten}\ \emph {et~al.}(2004)\citenamefont
  {Eichten}, \citenamefont {Lane},\ and\ \citenamefont {Quigg}}]{Eic04}%
  \BibitemOpen
  \bibfield  {author} {\bibinfo {author} {\bibfnamefont {E.~J.}\ \bibnamefont
  {Eichten}}, \bibinfo {author} {\bibfnamefont {K.}~\bibnamefont {Lane}}, \
  and\ \bibinfo {author} {\bibfnamefont {C.}~\bibnamefont {Quigg}},\ }\href
  {\doibase 10.1103/PhysRevD.69.094019} {\bibfield  {journal} {\bibinfo
  {journal} {Phys. Rev.D}\ }\textbf {\bibinfo {volume} {69}},\ \bibinfo {pages}
  {094019} (\bibinfo {year} {2004})},\ \Eprint
  {http://arxiv.org/abs/hep-ph/0401210} {arXiv:hep-ph/0401210 [hep-ph]}
  \BibitemShut {NoStop}%
\bibitem [{\citenamefont {Ferretti}\ \emph {et~al.}(2013)\citenamefont
  {Ferretti}, \citenamefont {Galatà},\ and\ \citenamefont
  {Santopinto}}]{Fer13}%
  \BibitemOpen
  \bibfield  {author} {\bibinfo {author} {\bibfnamefont {J.}~\bibnamefont
  {Ferretti}}, \bibinfo {author} {\bibfnamefont {G.}~\bibnamefont {Galatà}}, \
  and\ \bibinfo {author} {\bibfnamefont {E.}~\bibnamefont {Santopinto}},\
  }\href {\doibase 10.1103/PhysRevC.88.015207} {\bibfield  {journal} {\bibinfo
  {journal} {Phys. Rev.C}\ }\textbf {\bibinfo {volume} {88}},\ \bibinfo {pages}
  {015207} (\bibinfo {year} {2013})},\ \Eprint {http://arxiv.org/abs/1302.6857}
  {arXiv:1302.6857 [hep-ph]} \BibitemShut {NoStop}%
\bibitem [{\citenamefont {Ferretti}\ \emph {et~al.}(2014)\citenamefont
  {Ferretti}, \citenamefont {Galatà},\ and\ \citenamefont
  {Santopinto}}]{Fer14_5}%
  \BibitemOpen
  \bibfield  {author} {\bibinfo {author} {\bibfnamefont {J.}~\bibnamefont
  {Ferretti}}, \bibinfo {author} {\bibfnamefont {G.}~\bibnamefont {Galatà}}, \
  and\ \bibinfo {author} {\bibfnamefont {E.}~\bibnamefont {Santopinto}},\
  }\href {\doibase 10.1103/PhysRevD.90.054010} {\bibfield  {journal} {\bibinfo
  {journal} {Phys. Rev.D}\ }\textbf {\bibinfo {volume} {90}},\ \bibinfo {pages}
  {054010} (\bibinfo {year} {2014})},\ \Eprint {http://arxiv.org/abs/1401.4431}
  {arXiv:1401.4431 [nucl-th]} \BibitemShut {NoStop}%
\bibitem [{\citenamefont {Ferretti}\ and\ \citenamefont
  {Santopinto}(2014)}]{Fer14_9}%
  \BibitemOpen
  \bibfield  {author} {\bibinfo {author} {\bibfnamefont {J.}~\bibnamefont
  {Ferretti}}\ and\ \bibinfo {author} {\bibfnamefont {E.}~\bibnamefont
  {Santopinto}},\ }\href {\doibase 10.1103/PhysRevD.90.094022} {\bibfield
  {journal} {\bibinfo  {journal} {Phys. Rev.D}\ }\textbf {\bibinfo {volume}
  {90}},\ \bibinfo {pages} {094022} (\bibinfo {year} {2014})},\ \Eprint
  {http://arxiv.org/abs/1306.2874} {arXiv:1306.2874 [hep-ph]} \BibitemShut
  {NoStop}%
\bibitem [{\citenamefont {Eichten}\ \emph {et~al.}(2008)\citenamefont
  {Eichten}, \citenamefont {Godfrey}, \citenamefont {Mahlke},\ and\
  \citenamefont {Rosner}}]{Eic08}%
  \BibitemOpen
  \bibfield  {author} {\bibinfo {author} {\bibfnamefont {E.~J.}\ \bibnamefont
  {Eichten}}, \bibinfo {author} {\bibfnamefont {S.}~\bibnamefont {Godfrey}},
  \bibinfo {author} {\bibfnamefont {H.}~\bibnamefont {Mahlke}}, \ and\ \bibinfo
  {author} {\bibfnamefont {J.~L.}\ \bibnamefont {Rosner}},\ }\href {\doibase
  10.1103/RevModPhys.80.1161} {\bibfield  {journal} {\bibinfo  {journal} {Rev.
  Mod. Phys.}\ }\textbf {\bibinfo {volume} {80}},\ \bibinfo {pages} {1161}
  (\bibinfo {year} {2008})},\ \Eprint {http://arxiv.org/abs/hep-ph/0701208}
  {arXiv:hep-ph/0701208 [hep-ph]} \BibitemShut {NoStop}%
\bibitem [{\citenamefont {Segovia}\ \emph {et~al.}(2019)\citenamefont
  {Segovia}, \citenamefont {Steinbeißer},\ and\ \citenamefont
  {Vairo}}]{Vai19}%
  \BibitemOpen
  \bibfield  {author} {\bibinfo {author} {\bibfnamefont {J.}~\bibnamefont
  {Segovia}}, \bibinfo {author} {\bibfnamefont {S.}~\bibnamefont
  {Steinbeißer}}, \ and\ \bibinfo {author} {\bibfnamefont {A.}~\bibnamefont
  {Vairo}},\ }\href {\doibase 10.1103/PhysRevD.99.074011} {\bibfield  {journal}
  {\bibinfo  {journal} {Phys. Rev.D}\ }\textbf {\bibinfo {volume} {99}},\
  \bibinfo {pages} {074011} (\bibinfo {year} {2019})},\ \Eprint
  {http://arxiv.org/abs/1811.07590} {arXiv:1811.07590 [hep-ph]} \BibitemShut
  {NoStop}%
\bibitem [{\citenamefont {De~Fazio}(2009)}]{DeF08}%
  \BibitemOpen
  \bibfield  {author} {\bibinfo {author} {\bibfnamefont {F.}~\bibnamefont
  {De~Fazio}},\ }\href {\doibase 10.1103/PhysRevD.79.054015} {\bibfield
  {journal} {\bibinfo  {journal} {Phys. Rev.D}\ }\textbf {\bibinfo {volume}
  {79}},\ \bibinfo {pages} {054015} (\bibinfo {year} {2009})},\ \bibinfo {note}
  {[Erratum: Phys. Rev.D83,099901(2011)]},\ \Eprint
  {http://arxiv.org/abs/0812.0716} {arXiv:0812.0716 [hep-ph]} \BibitemShut
  {NoStop}%
\end{thebibliography}%

\end{document}